\documentclass[journal]{vgtc}                %

\ifpdf%
  \pdfoutput=1\relax                   %
  \usepackage{graphicx}                %
  \DeclareGraphicsExtensions{.pdf,.png,.jpg,.jpeg} %
\else%
  \usepackage{graphicx}                %
  \DeclareGraphicsExtensions{.eps}     %
\fi%

\graphicspath{{figures/}{pictures/}{images/}{./}} %

\usepackage{microtype}                 %
\PassOptionsToPackage{warn}{textcomp}  %
\usepackage{textcomp}                  %
\usepackage{mathptmx}                  %
\usepackage{times}                     %
\usepackage{cite}                      %
\usepackage{tabu}                      %
\usepackage{booktabs}                  %
\usepackage{tabularx}
\usepackage{enumitem}
\usepackage{xspace}
\usepackage{scalerel}

\usepackage[final]{review}

\usepackage{xr}

\makeatletter
\newcommand*{\addFileDependency}[1]{%
  \typeout{(#1)}
  \@addtofilelist{#1}
  \IfFileExists{#1}{}{\typeout{No file #1.}}
}
\makeatother
 
\newcommand*{\myexternaldocument}[1]{%
    \externaldocument{#1}%
    \addFileDependency{#1.tex}%
    \addFileDependency{#1.aux}%
}

\myexternaldocument{___supplementary-material}

\newcommand{\pilingjs}{\textsc{Piling.js}\xspace}

\newcommand{\tgrouping}{\textbf{T1}\xspace}
\newcommand{\tarrangement}{\textbf{T2}\xspace}
\newcommand{\tpreviewing}{\textbf{T3}\xspace}
\newcommand{\tbrowsing}{\textbf{T4}\xspace}
\newcommand{\taggregation}{\textbf{T5}\xspace}

\onlineid{0}

\vgtccategory{Research}
\vgtcpapertype{system}

\title{A Generic Framework and Library for\\Exploration of Small Multiples through Interactive Piling}

\author{Fritz Lekschas, Xinyi Zhou, Wei Chen, Nils Gehlenborg, Benjamin Bach, and Hanspeter Pfister}
\authorfooter{
  \item
    F. Lekschas and H. Pfister are with Harvard School of Engineering and Applied Sciences, Cambridge, MA, USA.\\
  \item
    X. Zhou and W. Chen are with State Key Lab of CAD\&CG, Zhejiang University, Hangzhou, China.\\
  \item
    B. Bach is with the University of Edinburgh, Edinburgh, UK.\\
  \item
    N. Gehlenborg is with Harvard Medical School, Boston, MA, USA.\\
}

\shortauthortitle{Lekschas \MakeLowercase{\textit{et al.}}: A General Framework and Library for Exploration of Small Multiples through Interactive Piling}

\abstract{
Small multiples are miniature representations of visual information used generically across many domains. Handling large numbers of small multiples imposes challenges on many analytic tasks like inspection, comparison, navigation, or annotation. To address these challenges, we developed a framework and implemented a library called \pilingjs for designing interactive \emph{piling} interfaces. Based on the piling metaphor, such interfaces afford flexible organization, exploration, and comparison of large numbers of small multiples by interactively aggregating visual objects into piles.
Based on a systematic analysis of previous work, we present a structured design space to guide the design of visual piling interfaces.
To enable designers to efficiently build their own visual piling interfaces, \pilingjs provides a declarative interface to avoid having to write low-level code and implements common aspects of the design space. An accompanying GUI additionally supports the dynamic configuration of the piling interface.
We demonstrate the expressiveness of \pilingjs with examples from machine learning, immunofluorescence microscopy, genomics, and public health.
}

\keywords{Information visualization, small multiples, interactive piling, visual aggregation, spatial organization.}

\teaser{
  \centering
  \includegraphics[width=\linewidth]{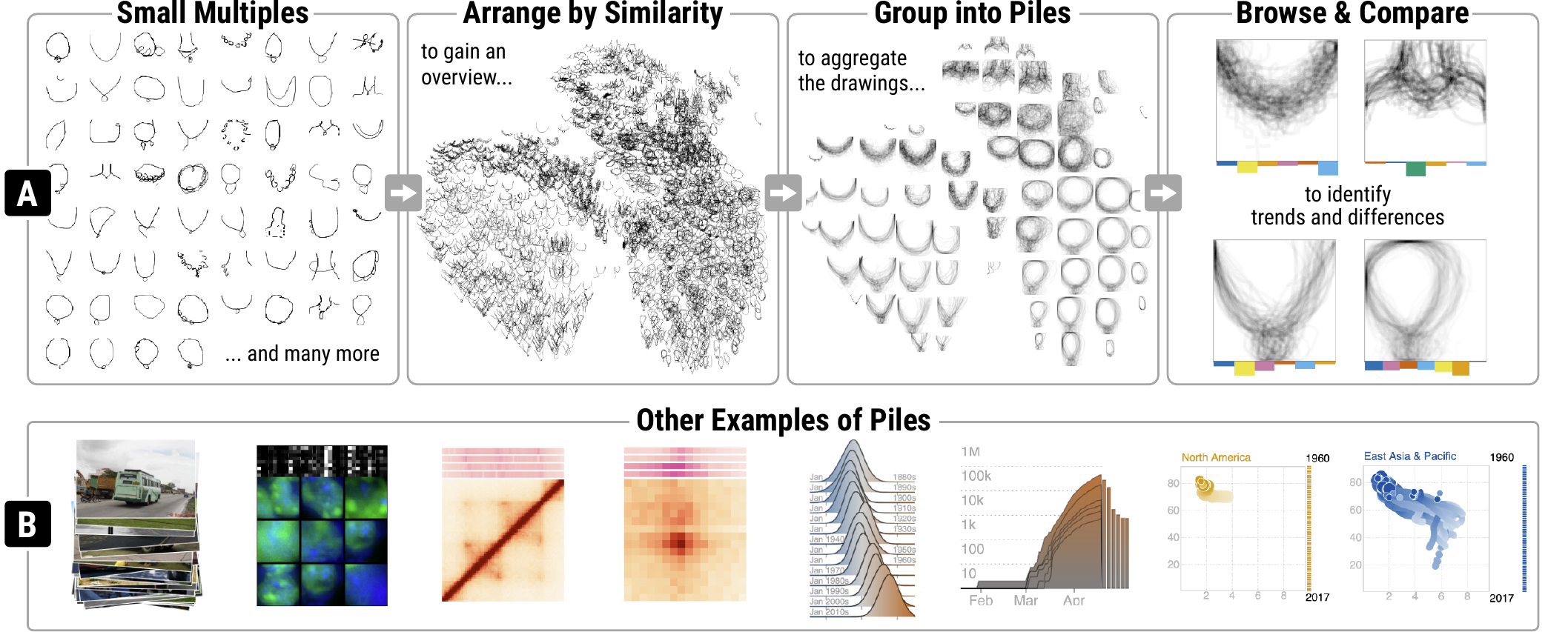}
  \caption{\textbf{Exploring Small Multiples through Visual Piling.} (A) An example of thousands of necklace sketches from Google Quickdraw~\cite{quickdraw} displayed as small multiples. The interactive arrangement, grouping, and aggregation of small multiples into piles support the discovery and comparison of reoccurring patterns. (B) Other types of small multiple visualizations grouped and aggregated into piles, including (from left to right) natural and immunofluorescence microscopy images, matrices, area charts, and scatterplots.}
  \label{fig:teaser}
}

\vgtcinsertpkg

\begin{document}

\firstsection{Introduction}

\maketitle

In many disciplines, datasets consist of large numbers of elements, pattern instances, or dimensions. For instance, in supervised machine learning, researchers compile sets of photos to train and validate machine learning models; in genomics, computational biologists study visual patterns that act as proxies for biological features; in public health, medical experts try to correlate different measurements to health conditions of their patient cohort.

Small multiples~\cite{tufte1990envisioning} are a widely used visualization technique to display such datasets through a series of miniature visualizations that show different facets or subsets of the data. However, as the number of small multiples grows, comparison and exploration can become inefficient due to the decreasing availability of screen real estate per visualization and the increasing efforts for sequential scanning. Sub-sampling or filtering can help to limit the number of small multiples but might obscure important characteristics of the dataset. Summary visualizations can alleviate this problem by aggregating subsets of the data into a single visualization. However, the analyst needs to know upfront how to organize the dataset into subsets. Without interactive features, exploration with summary visualizations can be limited when there are many potentially interesting facets or subsets to explore.

We propose a generic framework for exploring large numbers of small multiples through interactive visual \emph{piling}. Inspired by how physical piles enable casual organization~\cite{mander1992pile} of paper documents, piling in visualization affords spatial grouping of visual elements into piles that can be arranged, browsed, and aggregated interactively. By combining the benefits of small multiples and visual aggregations with interactive browsing, piling can be an effective technique for exploring small multiples. For instance, in \autoref{fig:teaser}A, we demonstrate how piling enables the discovery and comparison of shared concepts of necklace sketches through interactive arrangements, groupings, aggregation, and browsing.
Currently, piling has been applied to matrix visualizations by ad-hoc domain-specific methods to explore set typed data~\cite{sadana2014onset}, dynamic networks~\cite{bach2015small}, and matrix patterns~\cite{lekschas2018hipiler}. But there are many more scenarios where interactive visual piling can be useful for exploration, as shown in~\autoref{fig:teaser}. However, developing a new piling system for every use case would be time-consuming. Moreover, many concepts of piling are independent of the domain, data type, and visual encoding. 

We present the first overview of interactive visual piling as a method for the exploration of small multiples. Based on a systematic analysis of previous work, we define a design space according to five analytical tasks that any piling interface should support: grouping, arrangement, previewing, browsing, and aggregation. We focus on analytical tasks to study how different visual encoding and interaction approaches support the exploration of small multiples. Our design space provides guidance for the design of future piling interfaces. To streamline the implementation of piling interfaces, we developed \pilingjs---a JavaScript-based library that provides solutions for many common aspects of our design space. \pilingjs is built around a data-type independent rendering pipeline and a declarative view specification to avoid having to write low-level code for handling the interactive piling interface. Using \pilingjs, we demonstrate the generality of interactive piling for exploring small multiples and the expressiveness of our library with examples from machine learning, immunofluorescence microscopy, geography, public health, genomics, and more. The source code of \pilingjs is freely available under a permissive open source license at \url{https://github.com/flekschas/piling.js} and features extensive documentation and examples at \url{https://piling.js.org}.

\section{Related Work} \label{sec:related-work}

\textbf{Small Multiples.} Small multiples~\cite{tufte1990envisioning} are a series of miniature visualizations that show different facets or subsets of a dataset or different instances of a pattern type. Small multiples afford direct visual comparison and are, for example, used for multifaceted exploration~\cite{bavoil2005vistrails,googlefacets}, analyzing temporal data~\cite{robertson2008effectiveness,boyandin2012qualitative}, parameterization~\cite{marks1997design,jankun2001visualization}, or as a general exploration method for visual analytics~\cite{van2013small}.
Small multiple designs are conceptually similar to glyph design~\cite{ropinski2008taxonomy} as the reduced availability of screen real estate render glyphs useful.
Typically, small multiples are positioned in a grid or data-driven layout~\cite{ward2002taxonomy,smilkov2016embedding} that the user cannot manipulate directly. With visual piling, the goal is to enhance small multiples to support interactive grouping and aggregation of large numbers of small multiples for scenarios that require comparisons across a multitude of facets and subsets of the data.

\textbf{Piling for Document Organization.} Piling is a technique for the spatiotemporal organization of documents. Based on studying physical piling, Thomas Malone~\cite{malone1983people} suggests that piling is cognitively easier than filing as it only involves loose classification of documents.
In a study conducted by Whittaker and Hirschberg~\cite{whittaker2001character}, physical piling resulted in more frequently-browsed data collections compared to a folder-based exploration approach.
Mander et al.~\cite{mander1992pile} introduced piling as a technique for casual organization of documents in a virtual desktop environment. To address the scalability issues of piling, they experimented with automatic piling strategies and different modes of visually encoding and interacting with piles to aid document retrieval and browsing. In follow-up work~\cite{rose1993content}, they show how automatic piling-based grouping and aggregation can enhance content-aware browsing of virtual document collections. Kim et al.~\cite{kim2012compact} employed visual piling for browsing photos and found automatic piling to be as efficient for search as manually sorting and more efficient compared to automatically ordering. In this work, we are expanding the piling approach for casual document organization into a generic technique for interactive visual aggregation.

\textbf{Interacting with Visual Piles.} In the human computer interaction community, several projects explored interaction techniques for interactive browsing of document-based piles. For instance, using a 3D virtual desktop environment called BumpTop~\cite{agarawala2006keepin}, Agarawala et al. studied and implemented several pen-based interaction techniques for a tabletop display. BumpTop explores leafing through items like one would leaf through pages of a book, partially dispersing piles to see individual items better, or temporarily dispersing piles into a grid of items to avoid any overlap. Aliakseyeu et al.~\cite{aliakseyeu2007interacting} have further studied pen-based interaction techniques for browsing piles and found dispersing to be most effective. Other work explored the space of tangible interactions with piles in a mixed-reality environment such as digital tabletops~\cite{khalilbeigi2010interaction} or bendable e-ink displays~\cite{girouard2012displaystacks}.

Additionally, Bauer et al.~\cite{bauer2004computationally,bauer2005spatial} have explored spatial arrangement techniques for piling in Dynapad, where a pile is more loosely defined as a spatially-constrained set of visual items. Items must not necessarily overlap, which allows for continual exposure of items but requires more space and does not support aggregation. Another interaction technique for spatially organized items, called Bubble Clusters~\cite{watanabe2007bubble}, is built on the implicit formation of groups based on their spatial proximity. WallTop~\cite{bi2014walltop} implements a similar approach, where overlapping windows feature an outline that allows for group-based spatial positioning via drag-and-drop.
We analyzed this work to identify common gestures for interaction, which we implemented in \pilingjs.

\textbf{Piling for \new{Information} Visualizations.} In information visualization, visual piling is used for comparison. \new{Tominski et al.~\cite{tominski2012interaction} developed a generic interactive technique for pairwise comparison of information visualizations inspired by how people compare physical sheets of paper. Their technique is similar to piling for pairwise comparisons, but it does not visually or interactively scale to more than two items.}
Beyond this work, piling has mainly been applied to matrices for visual aggregation. For example, the Onset~\cite{sadana2014onset} technique implements a piling interface to interactively aggregate binary matrices for comparison.
Bach et al. extended this approach to dynamic networks in Small Multipiles~\cite{bach2015small} for detecting states over time. Importantly, Bach et al. introduced the notion of a \emph{preview} representation for items, which they implement as one-dimensional aggregates of the 2D matrix to aid browsing. In HiPiler~\cite{lekschas2018hipiler}, the idea of piling is further generalized to support one-, two-, and multi-dimensional arrangements. Vogogias et al.~\cite{vogogias2018bayespiles} and Fernandez et al.~\cite{fernandez2018domain} have applied similar matrix-based visual piling ideas to other applications in biology and software evolution.
Finally, Lekschas et al.~\cite{lekschas2019pattern} use piling to guide navigation in multiscale visualizations by aggregating overlapping patterns into piles and displaying them as insets.

In general, visual piling is an approach to reduce clutter~\cite{ellis2007taxonomy} through interactive aggregation.
In this paper we demonstrate the usefulness of piling and generalize the piling approach beyond matrix visualizations.

\section{The Visual Piling Approach} \label{sec:piling-approach}

Visual piling is an interactive approach for organizing, exploring, and comparing small multiples. Piling is centered around the act of spatially positioning items on top of each other, which together form a pile, and arranging these piles meaningfully to support effective comparison.
\subsection{Elements and Properties of a Visual Pile} \label{sec:piling-approach:elements-properties}

\begin{figure}[t]
    \centering
    \includegraphics[width=1\columnwidth]{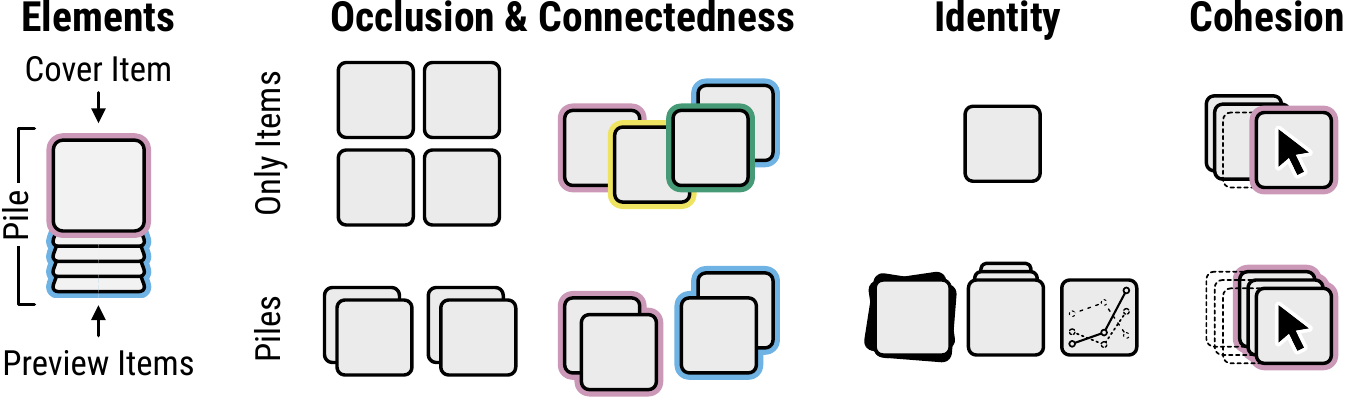}
    \caption{\textbf{Elements and Properties of Visual Piles.} To illustrate key properties of piles, we differentiate between individual items and piles.}
    \label{fig:pile-properties}
\end{figure}

Inspired by physical piles of paper documents, we define a pile as a group of partially-occluded small multiples that results from piling up individual \emph{items}, illustrated in \autoref{fig:pile-properties}. Given the partial overlap, only a single item is shown in its entirety, which we call the pile \emph{cover}. As the remaining items are only partially visible, we refer to them as \emph{previews}.
While there are many ways of visually representing a pile, we distinguish piles from other forms of spatially-arranged small multiples by the following set of properties, which builds upon the description of a physical pile from Bauer et al.~\cite{bauer2005spatial}.

\smallskip
\textbf{Occlusion \& Connectedness.} Items that comprise a pile should occlude each other partially to form a single mutually-connected unit. However, piles can temporarily be dispersed for exploration. %

\textbf{Identity.} A pile must differentiate itself visually from a single item. There are different visual encodings to identify a pile, like a label indicating the number of items on a pile, superimposed semi-transparent images, or items that are offset relative to each other.

\textbf{Cohesion.} Items on a pile should act as a single element during the exploration. Cohesive behavior does not mean that access to individual items is lost upon grouping items into a pile. However, a pile should reflect the notion of a group when interacting with the piling interface.

\textbf{\new{Transience}.} Piling should be seen as a dynamic process \new{where the piling state can change frequently}. In contrast to other aggregation techniques, the goal of piling is to compose and disperse piles interactively rather than to just consume a static grouping state. \new{However, this does not mean that piles cannot persist.}

\subsection{Goals and Tasks} \label{sec:piling-approach:goals-tasks}

Even though the application-specific goals differ, we identify two overarching goals for interactive visual piling interfaces \new{from related work}. (\textbf{G1}) Visual piling is a tool for organizing data collections into subsets to reduce complexity. This includes, for example, to sort items into groups, categorize groups based on their content, or filter out subsets of items for quality control. (\textbf{G2}) Beyond organization, visual piles are a means to explore and compare individual items and groups of items to each other. Specifically, one might want to determine the primary topic of a group, identify outliers, or discover trends.

To identify the common tasks needed to support organization, exploration, and comparison, we systematically reviewed related work. Following an open-coding approach, the first two authors coded all 17 piling-related papers from~\autoref{sec:related-work} according to their application-specific tasks independently. We focused our coding efforts on the role of interactive piling to not confuse piling-specific with unrelated tasks. After coding the papers, the first two authors resolved disagreements. Subsequently, we generalized the assigned codes into five high-level analytic tasks that any interactive visual piling interface should support.

\begin{itemize}
\setlength\itemsep{0.01em}

\item [\textbf{T1}] \textbf{Grouping:} \emph{manually} or \emph{automatically} sort items into piles.

\item [\textbf{T2}] \textbf{Arrangement:} position items and piles relative to each other in an \emph{orderly}, \emph{randomized}, \emph{gridded}, or \emph{unconstrained} layout.

\item [\textbf{T3}] \textbf{Previewing:} identify and locate items on a pile using \emph{in-place}, \emph{gallery}, \emph{foreshortened}, \emph{combining}, and \emph{indicating} previews.

\item [\textbf{T4}] \textbf{Browsing:} search, explore, and navigate within and between piles through \emph{in-place}, \emph{dispersive}, \emph{layered}, and \emph{hierarchical} browsing.

\item [\textbf{T5}] \textbf{Aggregation:} summarize a pile into a \emph{synthesized}, \emph{representative}, or \emph{abstract} representation.

\end{itemize}

To study how different visual encoding and interaction approaches support the exploration of small multiples, we use these five analytical tasks to structure the design space exploration (\autoref{sec:design-space}) and to guide future piling designs.

\subsection{Usage Scenario} \label{sec:piling-approach:usage-scenario}

To exemplify how visual piling enhances the exploration of small multiples, we describe a typical usage scenario following the example of necklace sketches from Google Quickdraw~\cite{quickdraw} (\autoref{fig:teaser}A), which we also demonstrate in the supplementary video.
One goal in analyzing large collections of visual objects is to identify and compare trends within the dataset. Inspired by Forma Fluens~\cite{formafluens}, we are trying to find reoccurring pattern concepts. Visualizing the sketches as small multiples (\autoref{fig:teaser}A1 left) allows us to assess and compare individual sketches, but it does not support the discovery of shared concepts.
A common approach to uncover similarities within large and high-dimensional data collections is to arrange (\tarrangement) the items as a two-dimensional embedding for cluster analysis (\autoref{fig:teaser}A2). We arranged the items by a UMAP~\cite{mcinnes2018umap} embedding of image features that were learned with a convolutional autoencoder. In the resulting cluster plot, items can be represented as a symbol (e.g., a dot) or a small thumbnail. While symbol-based cluster plots are highly scalable, they do not reveal the visual details of a cluster. On the other hand, thumbnail-based cluster plots do not scale to large datasets due to overplotting issues.
Visual piling provides a trade-off by grouping (\tgrouping) spatial clusters, i.e., clusters formed by items in relative proximity (\autoref{fig:teaser}A3).
By aggregating (\taggregation) all sketches into an average and showing this average as the pile cover, we can discover and \textit{browse} overarching concepts effectively. For instance, \new{after manually refining the grouping and arrangement of four piles (Supplementary Figure S1), we can see that} people are sketching a necklace as an open beaded necklace, a necklace worn around a neck, an open pendant necklace, or a closed pendant necklace (\autoref{fig:teaser}A4). Visual piling also affords the encoding of additional information beyond the individual items. For instance, in \autoref{fig:teaser}A4, we visualize the relative distribution of geographic regions across a pile using small bar charts below each pile.

\section{A Design Space for Visual Piling} \label{sec:design-space}

\begin{figure*}
    \centering
    \includegraphics[width=1\textwidth]{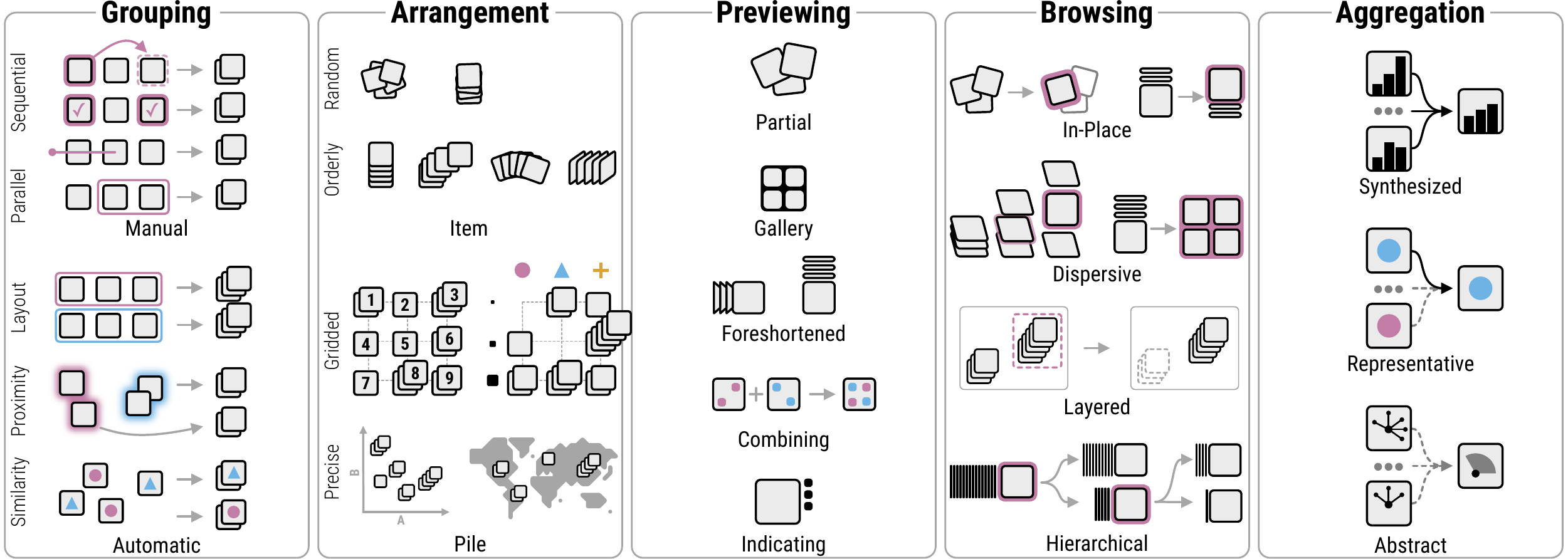}
    \caption{\textbf{Dimensions of the Visual Piling Design Space.} We structure the design space according to the five analytical tasks into different approaches for grouping, arrangement, previewing, browsing, and aggregation.}
    \label{fig:dimensions}
\end{figure*}

This is the first design space (\autoref{fig:dimensions}) for visual piling. For each of the five analytical tasks (\autoref{sec:piling-approach:goals-tasks}), we derived general approaches and common solutions from previous work through multiple discussions among the co-authors. The resulting subcategories cover overarching approaches of each task. We generalize these approaches to highlight conceptual differences. Multiple approaches can be combined to offer different ways of organizing and exploring small multiples. In our design space, we cover the relevant visual encodings and interactions. \new{We also describe common gestures for triggering interactions but do not attempt to provide a complete overview of all possible gestures.}

\subsection{Grouping} \label{sec:design-space:grouping}

We distinguish between \textit{manual} and \textit{automatic} grouping (\tgrouping), as exemplified in~\autoref{fig:dimensions} Grouping. Manual grouping requires the user to interactively determine which items should be grouped and, potentially, in which order. Automatic grouping follows a specific procedure to group multiple items at once.

\smallskip
\textbf{Manual.}
\textit{Sequential} grouping is the simplest form of manual grouping. It requires the user to group items interactively, one at a time. This is typically achieved with a drag-and-drop gesture~\cite{mander1992pile,bauer2004computationally,agarawala2006keepin,watanabe2007bubble,sadana2014onset,bach2015small,lekschas2018hipiler}. While sequential grouping requires more time, it enables temporal organization. For instance, the most recently added elements can be located on top of the pile.
For efficiency, one can also form a group from multiple selected items. While multi-select grouping does not result in intermediate groupings, the sequence of selected items can still be reflected, given the order of selected items.
In contrast, \textit{parallel} grouping techniques allow two or more items to be piled up at the same time. For instance, many piling interfaces support region-based grouping via lasso techniques~\cite{agarawala2006keepin,lekschas2018hipiler}.
Parallel grouping does not afford temporal organization as the order in which multiple items are grouped together is not \new{explicitly} defined.
A special form of grouping, \new{which can be treated as parallel or sequential,} is swiping~\cite{lekschas2018hipiler}, where the user moves the mouse cursor or pen over each item to be grouped. Swiping enables more precise selections in dense arrangements like cluster plots.

\smallskip
\textbf{Automatic.}
Many piling interfaces support automatic grouping to improve scalability.
\textit{Layout}-driven grouping is based on an explicitly- or implicitly-defined layout. Items that are located within the same unit of the layout can then be grouped. Such units can, for instance, be the rows, columns, or grid cells~\cite{lekschas2018hipiler}.
\textit{Proximity}-based grouping uses the Gestalt principle of ``proximity,'' which states that nearby items implicitly form groups perceptually.
However, implicit grouping can cause uncertainty in subsequent pile interactions~\cite{mander1992pile} as it is not always possible to infer the grouping state as perceived by the user~\cite{bauer2005spatial}. Therefore, most piling interfaces only use proximity to trigger explicit grouping, e.g., by outlining the pile bounds~\cite{bauer2004computationally,watanabe2007bubble} or merging nearby items~\cite{lekschas2019pattern}.
Finally, in \textit{similarity}-based grouping, items are merged automatically based on some notion of similarity. While there are many different ways of measuring similarity, fundamentally, the similarity can be derived from the items~\cite{rose1993content,kim2012compact,bach2015small,lekschas2018hipiler} or related metadata~\cite{lekschas2018hipiler}.

\subsection{Arrangement} \label{sec:design-space:arrangement}

For arrangements (\tarrangement), we consider the relative positioning of items on a pile and the absolute positioning of piles (\autoref{fig:dimensions} Arrangement).
\smallskip
\textbf{Item arrangement.}
A \textit{random} item arrangement is characterized by non-deterministic offsets and rotations of the items. Such arrangements make the visual pile closely resemble a physical pile, which can be useful to distinguish between automatically- and manually-composed piles~\cite{mander1992pile}. Random item arrangements can also encode access patterns, such as the frequency of file access~\cite{digioia2005social}. Finally, a pseudo-random item arrangement can be the result of sequential grouping. Since it is unlikely that the user will stack items in a pixel-precise manner, the resulting offset can appear random. Nevertheless, the offset can provide meaningful cues to the pile creator~\cite{malone1983people} for browsing (\tbrowsing). 
In contrast, \textit{orderly} item arrangements are the result of automatic and deterministic positioning. Such arrangements enable controlling how much each item is overlapped, which is useful for comparing items~\cite{bach2015small,lekschas2018hipiler}. When the item offset follows a single direction, \textit{in-place} browsing can be efficient as the cursor movement is minimal. Also, orderly arrangements typically follow the Gestalt principle of ``continuation'' to foster perceptual grouping.

\smallskip
\textbf{Pile Arrangement.}
Dividing the canvas into rows and columns of a specific size leads to a \textit{gridded} pile arrangement. Gridded arrangements are useful for comparing piles due to the alignment. Imposing a specific ordering onto the pile can highlight temporal or sequential patterns. The simplest form of a gridded pile arrangement is a one-dimensional timeline~\cite{bauer2005spatial,kim2012compact} but two-dimensional grid layouts are more common to make use of the entire screen~\cite{bauer2004computationally,agarawala2006keepin,bach2015small,lekschas2018hipiler}.
Finally, as \textit{\new{precise}} pile arrangements, we summarize manual, layout-driven, or data-driven arrangements that require a pixel-precise positioning on the canvas. In this regards, automatic arrangements can incorporate one-dimensional~\cite{bauer2005spatial}, two-dimensional~\cite{lekschas2018hipiler}, or multidimensional~\cite{lekschas2018hipiler} scatterplots. The position can also be inherent to the items themselves, which is, for example, the case for exploring annotated pattern instances~\cite{lekschas2019pattern}.

\subsection{Previewing} \label{sec:design-space:previewing}

To afford content-awareness, visual piles can implement different \new{layout types to support} previewing items (\tpreviewing), as shown in~\autoref{fig:dimensions}, which is key to support effective exploration and navigation (\tbrowsing).

\textbf{Partial.}
Inspired by physical piling, partial previewing of items arises naturally and is implemented in many piling interfaces~\cite{bauer2004computationally,agarawala2006keepin,aliakseyeu2007interacting,khalilbeigi2010interaction,kim2012compact}. The effectiveness depends on the data type and the size of the partial previews.

\textbf{Gallery.}
When the partial overlap severely limits the perception of the item's content, one can opt for a gallery preview where a small number of items or aggregates~\cite{lekschas2019pattern} is shown in a regular grid. This approach can be useful for datasets in combination with a representative aggregation approach (\autoref{sec:design-space:aggregation}).

\textbf{Foreshortened.}
To limit the size of a pile while still providing item-specific previews, previews can be foreshortened along one axis. Such previews can be implemented with perspective distortion~\cite{agarawala2006keepin}, compression, or aggregation along one dimension~\cite{bach2015small,lekschas2018hipiler,fernandez2018domain,vogogias2018bayespiles,lekschas2019pattern}. While the first option can provide cues for search and navigation (\tbrowsing), the latter enables more effective comparison between alignable items.

\textbf{Combining.}
When working with items that have a sparse visual representation and shared axes, like scatterplots, line charts, or bar charts, multiple items can be combined to provide an overview. A combined preview can be the result of superimposing items with a transparent background or through dedicated aggregation~\cite{sadana2014onset}. While this approach is space-efficient, the relationship between the overview and individual items might get lost without employing other means of previewing items.

\textbf{Indicating.}
To maximize space efficiency while still hinting at an item's content, one can provide abstract indicators as item previews. Most common indicators are implemented as tabs~\cite{jakobsen2010piles,sadana2014onset}, but the indicator can be more abstract, e.g., a small dot. While indicators do not directly preview the content, they afford browsing (\autoref{sec:design-space:browsing}) and can encode metadata like the distribution of items.

\subsection{Browsing} \label{sec:design-space:browsing}

In the context of visual piling, we regard browsing (\tbrowsing) as the act of inspecting the visual details of piled items (\autoref{fig:dimensions} Browsing). Since browsing requires interaction, the applicability of different browsing approaches depends on the type of preview.

\textbf{In-Place.}
Inspecting the visual details of an item in-place is a fast browsing approach as the arrangement of items remains unchanged. To show an item in its entirety, the ordering of items is altered temporarily such that the browsed item is shown on top~\cite{agarawala2006keepin,aliakseyeu2007interacting,jakobsen2010piles}.
A variation of in-place browsing shows the browsed item next to the pile as an inset~\cite{mander1992pile}.
Another in-place browsing technique called ``leafing''~\cite{agarawala2006keepin,bach2015small,lekschas2018hipiler,lekschas2019pattern} involves interacting with foreshortened previews. Upon interaction, foreshortened previews \new{can either be} expanded to their full extent \new{or shown on top of the pile}, similar to flipping through the pages of a book.
Typically, in-place browsing is triggered by moving the pointing device over the preview item to be shown in its full extent~\cite{mander1992pile,agarawala2006keepin,aliakseyeu2007interacting,jakobsen2010piles,bach2015small,lekschas2018hipiler,lekschas2019pattern}.

\textbf{Dispersive.}
Dispersive pile browsing techniques temporarily disperse a pile such that the overlap between items is resolved partially or entirely, allowing for subsequent comparison of the items. To aid maintaining a mental map of the items on a pile, many dispersive techniques use the same type of layout for positioning the dispersed items and only increase the spacing between items~\cite{agarawala2006keepin,watanabe2007bubble,bach2015small}. A more disruptive approach arranges the dispersed items into a regular grid~\cite{bauer2005spatial,agarawala2006keepin,aliakseyeu2007interacting}. Pile dispersion is often triggered by a double click or tap~\cite{aliakseyeu2007interacting,watanabe2007bubble,bach2015small,lekschas2018hipiler}, but many gestures have been explored too including horizontally moving a pointer device back and forth~\cite{mander1992pile}, hovering over a pile~\cite{agarawala2006keepin}, and context-menu induced dragging gestures~\cite{agarawala2006keepin}. 
Finally, an indirect way of dispersive browsing employs a zoom gesture in combination with automatic proximity-based grouping~\cite{bauer2004computationally,lekschas2019pattern}. Thereby, piles gradually disperse as the user zooms into a specific region.

\textbf{Layered.}
Increasing numbers of small multiples limit the available space for visual browsing. Layered browsing techniques temporarily hide other items and, thus, give more space to the browsed piles. Layering can be combined with dispersive browsing to support flexible pile exploration and sub-piling~\cite{lekschas2018hipiler}.

\textbf{Hierarchical.}
Finally, for piles of many items, it can be ineffective to browse all items at once. Instead, hierarchical clustering can be employed to enable hierarchical browsing such that the pile only disperses into a subset of piles from the next hierarchical level.

\subsection{Aggregation} \label{sec:design-space:aggregation}

Aggregation (\taggregation) is the act of summarizing piled items into a concise form (\autoref{fig:dimensions} Aggregation). The goal of aggregation is to improve the content awareness \new{when previewing a pile} and aid comparison between groups of items. \new{Therefore, the choice of the aggregation method depends on the layout type for previewing.}

\textbf{Synthesized.}
We call aggregation techniques that create a single image from a group of items synthesized aggregations. Hereby, the resolution or granularity of the aggregate is identical to the items. Summary statistics are commonly used for synthesized aggregations~\cite{sadana2014onset,bach2015small,lekschas2018hipiler,vogogias2018bayespiles,lekschas2019pattern} but other methods are possible. When the aggregate presents new or unseen information, it is useful to provide some means of previewing individual items~\cite{sadana2014onset,bach2015small,lekschas2018hipiler} to enable item-specific comparisons.

\textbf{Representative.}
For data types where individual items do not align well, such as natural images, synthesized aggregations are typically ineffective. Instead, the pile can be summarized by a single or multiple representative items which are typically visualized as gallery previews~\cite{lekschas2019pattern}. Through careful sampling, the selection of representative items can provide enough information to inform the user about a pile's primary content.

\textbf{Abstract.}
Finally, for non-alignable but well-defined data, one can employ abstract aggregation techniques. The goal of such techniques is to provide a simplistic or schematic representation of the pile's content where the resolution or granularity is reduced compared to the items. Simplistic aggregations provide limited content awareness. However, the aggregation can, nevertheless, hint at the category or type of items on a pile, which can be useful for navigation (\tbrowsing).

\subsection{Additional Pile Encodings} \label{sec:design-space:additional-encodings}

\begin{figure}
    \centering
    \includegraphics[width=1\columnwidth]{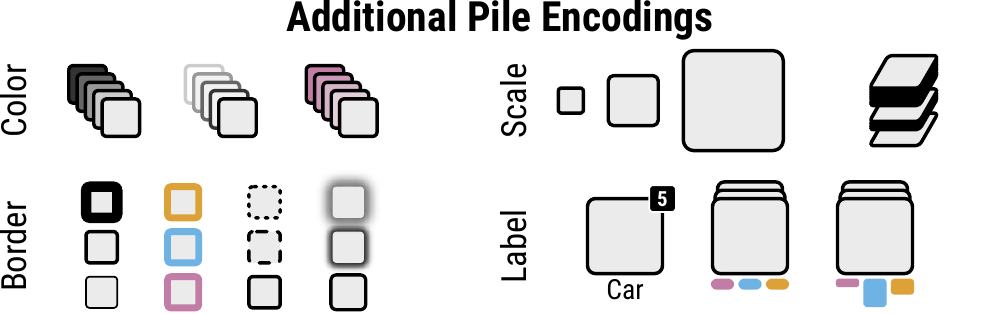}
    \caption{\textbf{Additional Pile Encoding.} The coloring, scale, border, and labels afford additional encodings that can be useful for exploration.}
    \label{fig:additional-encodings}
\end{figure}

Several additional style properties \new{(\autoref{fig:additional-encodings})} can be employed to encode related information such as the pile or item size, item access patterns, or categorical information.

\textbf{Coloring.}
To highlight trends within a group of piled items, one can adjust the lightness~\cite{mander1992pile} or apply other color filters. While potentially effective at encoding additional information, extreme color adjustments can harm content awareness and should be used with caution. 

\textbf{Scaling.}
Scaling the visual pile size in the x,y~\cite{lekschas2019pattern} or z-direction is another approach to encode additional information. Z-scaling requires that items are represented as three-dimensional objects. If applied, z-scaling can also afford edge browsing, which is an in-place browsing technique for physical piles~\cite{mander1992pile}.

\textbf{Border.}
Beyond manipulating the piled items, visual piles can also utilize a border to encode information through its color, size, texture, or sharpness~\cite{lekschas2018hipiler,lekschas2019pattern}.

\textbf{Labeling.}
When working with categorical data, grouping and comparison can be improved by drawing a textual label~\cite{mander1992pile,digioia2005social,bach2015small} or visualizing categories as colored badges~\cite{lekschas2018hipiler}. These badges can be adjusted to show the distribution of categories across a pile for improved browsing and comparison\new{, as shown in~\autoref{fig:teaser}A ``Browse \& Compare''}.

\section{\pilingjs -- A Library for Visual Piling} \label{sec:pilingjs}

Based on our design space definition, we developed \pilingjs, a JavaScript library, to streamline the implementation of piling interfaces. \pilingjs is built around a data-agnostic rendering pipeline and a declarative view specification to define complex piling interface without having to write low-level code (Supplementary Figure S2). The library manages the view state and implements methods for interactive grouping, arrangement, browsing, and \new{previewing}. The designer's task is to specify the data rendering and aggregation (i.e., the elements of a visual pile), and the visual pile encoding. \new{For interaction,} gestures to manipulate piles are provided automatically by \pilingjs. \new{See Supplementary Table S1 for an overview of \pilingjs' coverage of the visual piling design space.}

\subsection{Data Rendering} \label{sec:pilingjs:pipeline}

\begin{figure}[h]
    \centering
    \includegraphics[width=1\columnwidth]{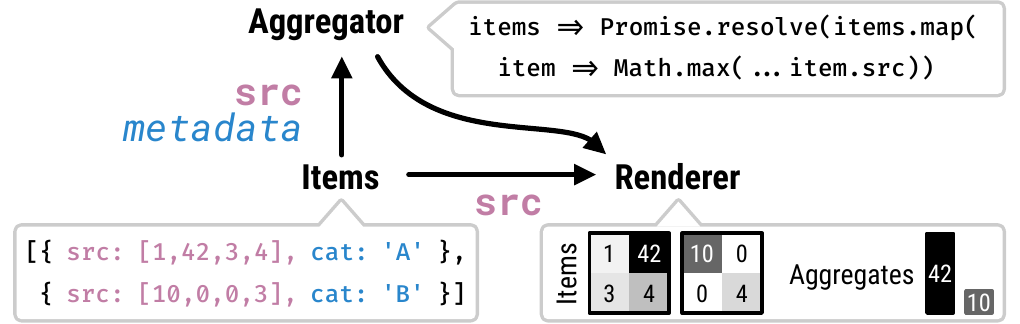}
    \caption{\textbf{Rendering Pipeline.} Renderers receive as input the \texttt{src} property of \texttt{items} or the result of an aggregator. The aggregator in turn receives the \texttt{items} as input including additional properties (blue).}
    \label{fig:rendering-pipeline}
\end{figure}

As a first step, the designer needs to specify the data rendering pipeline (\autoref{fig:rendering-pipeline}) to create the elements of a visual pile. This pipeline involves the definition of the item data, a renderer, and an aggregator. While all three aspects play together, their implementations are decoupled so that each component can be reused, replaced, and extended easily.

\smallskip
\textbf{Data.}
\pilingjs operates on two primary data concepts: \texttt{items} and \texttt{piles}. The \texttt{items} are determined by the user and treated as immutable objects, while \texttt{piles} are created dynamically by \pilingjs. Each item is expected to be a JavaScript object with a \texttt{src} property that is, by convention, passed to the renderer (\autoref{fig:rendering-pipeline}). This convention allows the designer to switch between different renderers without having to change the item itself. To enable exploration of dynamic datasets, \pilingjs supports a data matching strategy, using an optional \texttt{id} property, which works similarly to D3's~\cite{bostock2011d3} data joins. Upon updating the \texttt{items}, \pilingjs will match items by their \texttt{id} to identify outdated, updated, and newly-added items. By default, the \texttt{id} is set to the list index of \texttt{items}. An item object can contain additional properties to be used for dynamic styling, which enables concise and readable style specifications (\autoref{fig:dynamic-properties}).
The \texttt{piles} are a list of objects that hold the item \texttt{id}s and the pile's xy position. All other pile encoding aspects are inferred automatically from the view specification. We chose this relatively simple data model to make piling more accessible.

\begin{figure}[t]
    \centering
    \includegraphics[width=1\columnwidth]{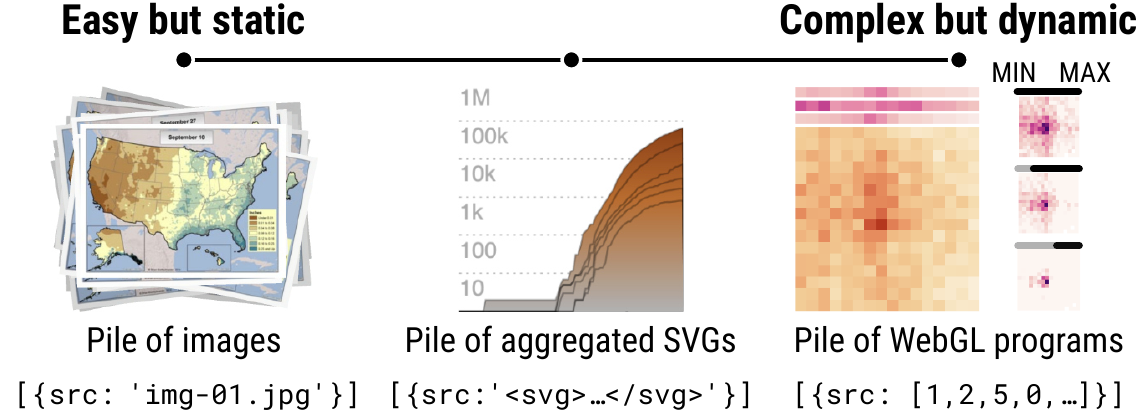}
    \caption{\textbf{Rendering Regimes.} Static renderers are easy to set up but offer limited support for aggregation. Dynamic renderers can be complex but support dynamic updates (e.g., color scaling) and aggregation.}
    \label{fig:rendering-regimes}
\end{figure}

\begin{figure}[b]
    \centering
    \includegraphics[width=0.95\columnwidth]{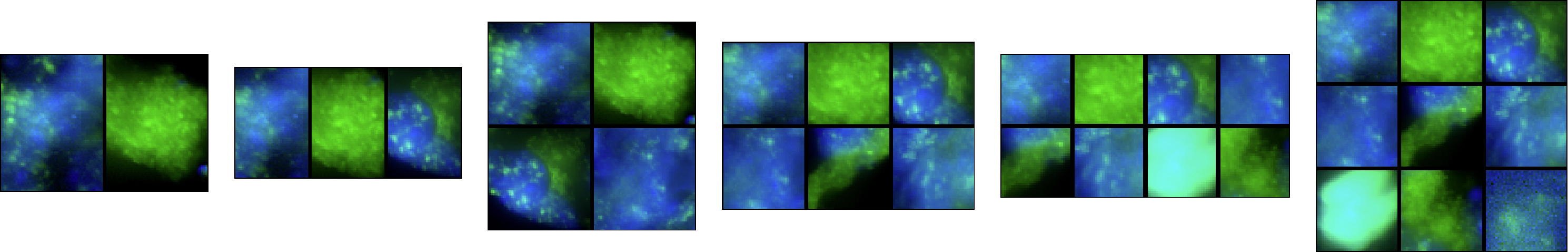}
    \caption{\textbf{Gallery Previews.} The built-in representative renderer supports galleries of two, three, four, six, eight, and nine previews.}
    \label{fig:representative-renderer}
\end{figure}

\smallskip
\textbf{Renderers.}
\pilingjs renders items into textures and subsequently operates on these textures during gesturing to decouple the state management and gesture handling from the domain-specific visualizations. To support a wide range of previewing and aggregation approaches, \pilingjs offers different rendering regimes for items, previews, and covers (\autoref{fig:rendering-regimes}).
A renderer is a function that receives as input the \texttt{src} property of one or more items and returns a single \textit{promise} object that resolves to texture resources once the rendering finishes (\autoref{fig:rendering-pipeline}). A promise is a proxy for a future value and commonly used to enable \textit{asynchronous} executions in JavaScript. The texture resources must be one of the following media types: an image, canvas, or video element. Since many web-based visualizations render SVGs, \pilingjs provides a predefined SVG renderer that accepts an SVG string or element as input. Using the SVG renderer, it is easy to render any static D3 visualization in \pilingjs. \pilingjs also includes a predefined matrix renderer and supports PixiJS~\cite{pixijs} WebGL programs as a renderer for dynamic re-rendering (\autoref{fig:rendering-regimes} right).
To support \textit{gallery} previews, \pilingjs implements \textit{meta} renderers, which compose multiple images into a single image. These renderers rely on a representative aggregation approach. For convenience, \pilingjs provides a built-in renderer that composes multiple items into a gallery (~\autoref{fig:representative-renderer}).
In general, each built-in renderer can be replaced or configured for customization.

\textbf{Aggregators.}
To support foreshortened item previewing or pile aggregations, the designer needs to specify an aggregator function for items or piles. The aggregator either receives as input a single item or multiple items and returns a single data source that is subsequently passed to the related renderer.
By decoupling renderers and aggregators, both can be reused without having to adjust them. Also, not all types of item previews require aggregation. For synthesizing aggregation, \pilingjs provides a set of predefined matrix aggregators that supports common summary statistics (mean, variance, and standard deviation). For representative aggregation, \pilingjs implements a generic cluster-based approach that employs k-means clustering as the computation is fast enough to not cause noticeable delays. We pick the $k$ items that are closest to the $k$ centroids from k-means as the representative images.

\subsection{Pile Encoding via View Properties}
\begin{figure}[t]
    \centering
    \includegraphics[width=1\columnwidth]{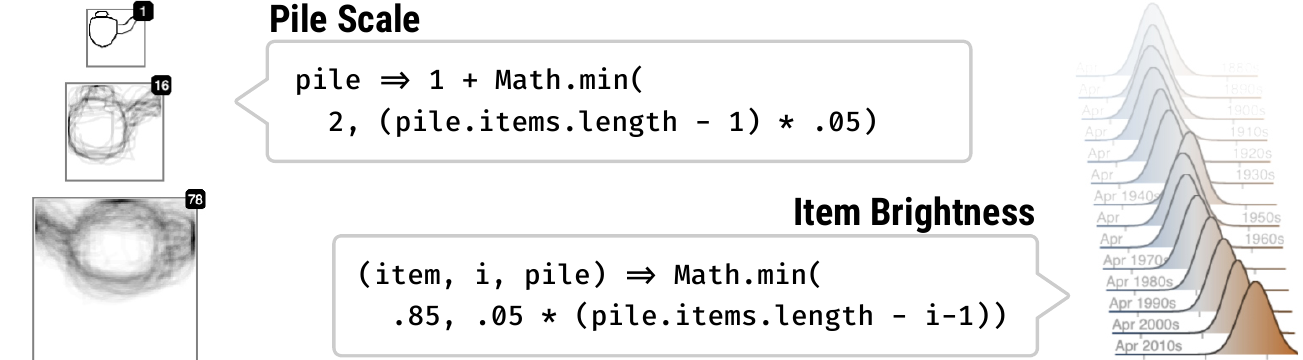}
    \caption{\textbf{Dynamic View Properties.} (Left) We dynamically scale the pixel size of the pile by the number of items. (Right) We dynamically adjust the brightness of items, based on the item order. Items with a small index appear brighter. The brightness ranges from -1 (black) to 1 (white).}
    \label{fig:dynamic-properties}
\end{figure}

\pilingjs offers many view properties to specify the arrangement, previewing, and visual encoding of piles (Supplementary Figure S2). For a complete overview of all view properties, please refer to \url{https://piling.js.org}. View properties are set via \pilingjs's \texttt{set} method, which receives as input the property name and value. In general, there are three types of view properties in \pilingjs, which related to our data model (\autoref{sec:pilingjs:pipeline}): \textit{global}, \textit{pile}-specific, and \textit{item}-specific view properties. Global properties relate to the entire piling interface and do not change during interactive grouping. For instance, \texttt{set('columns',10)} sets the number of columns to 10. In contrast, pile- and item-specific properties can depend on the grouping on the \texttt{pile}, i.e., the state of grouping, and therefore support dynamic specifier functions (\autoref{fig:dynamic-properties}). For instance, the pile border size could be a function of the number of elements on a pile. Inspired by D3~\cite{bostock2011d3}, \pilingjs implements a declarative data-driven interface to dynamic properties by passing the specific \texttt{items} and \texttt{piles} to the specifier function. Using this approach, the designer only needs to specify how to translate the data object into a property value, while \pilingjs visually renders the property. Dynamic pile-specific properties are invoked for every pile. They receive the current pile object and return a corresponding property value.
Item-specific property specifiers are invoked for every item on a pile as the pile's composition changes. The specifier function receives an item's data object, index, and corresponding pile object, and returns the property value (\autoref{fig:dynamic-properties} bottom-right).

\subsection{Pile Interactions}
\label{sec:pilingjs:interactions}

\new{While visual piling is agnostic to the input device (e.g., mouse, pen, or touch), \pilingjs currently focuses on mouse interactions} and implements general mouse gestures found across several related works (\autoref{sec:design-space}) for manual grouping, arrangement, browsing, as well as methods for automatic grouping and arrangement.

\begin{figure}[b]
    \centering
    \includegraphics[width=1\columnwidth]{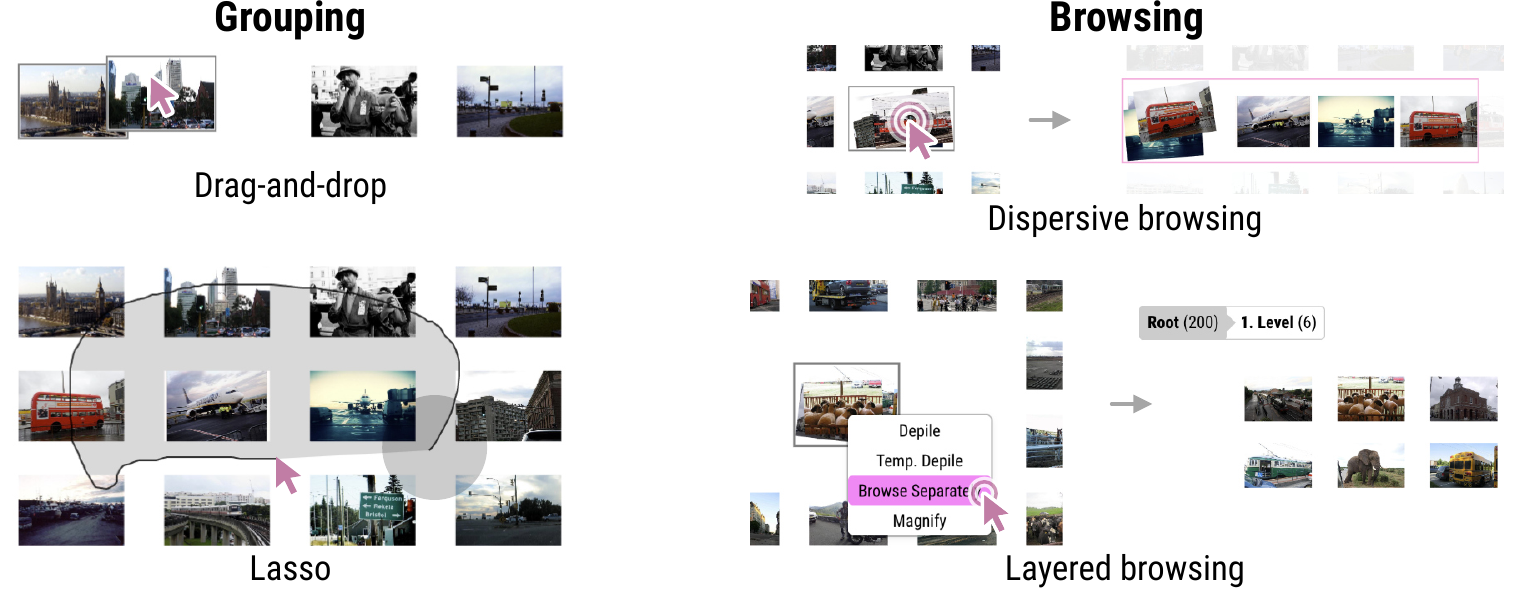}
    \caption{\textbf{Interaction Techniques.} \pilingjs supports drag-and-drop and lasso gestures for manual grouping. In \pilingjs, piles can be dispersed temporarily via a double click and browsed in layers via the context menu.}
    \label{fig:interactions}
\end{figure}

\smallskip
\textbf{Gestures for Manual Interactions.}
\pilingjs implements common gestures for grouping and arrangement (\autoref{sec:design-space}). Piles can be arranged manually via a drag-and-drop gesture. For sequential grouping, an item or pile needs to be dropped onto another item or pile (\autoref{fig:interactions} top-left). To group multiple items at once, \pilingjs offers a lasso tool. The lasso is initiated by clicking into an empty region of the canvas. Subsequently, a circle will appear (\autoref{fig:interactions} bottom-left). By clicking into this circle and holding down the primary mouse button, the user can start the lasso selection. All items located within the lasso area are grouped upon releasing the primary mouse button.
For browsing (\autoref{sec:design-space:browsing}), \pilingjs implements gestures for in-place, dispersive, layered, and hierarchical browsing. In-place browsing is triggered by a click on a pile and moving the mouse cursor over the previews. Double-clicking on a pile will temporarily disperse a pile into a regular grid, as shown in~\autoref{fig:interactions} (top-right). To browse a pile in layers, the user can activate the pile's context menu (\autoref{fig:interactions} top-right) and select ``browse separately.'' The browsed pile is additionally dispersed on the next layer to support rapid sub-piling.

\smallskip
\textbf{Automatic Grouping and Arrangements}
To support automatic grouping and arrangement, \pilingjs features a \texttt{groupBy}, \texttt{splitBy} and \texttt{arrangeBy} method. The \texttt{groupBy} method enables layout-, proximity-, and data-driven groupings, as described in~\autoref{sec:design-space:grouping}. As a complement, the \texttt{splitBy} method can split piles based on their position and data properties of items. Finally, the \texttt{arrangeBy} method enables automatic pile-specific arrangements.
All three methods rely on a \texttt{type} and an \texttt{objective}. The \texttt{type} determines the subroutine to be used and the \texttt{objective} provides the necessary data to execute this subroutine. For an example, see Supplementary Figure S3. Currently, \texttt{groupBy} supports proximity-based (\texttt{distance} and \texttt{overlap}), layout-driven (\texttt{grid}, \texttt{column}, and \texttt{row}), and similarity-based (\texttt{category} and \texttt{cluster}) grouping. For instance, \texttt{groupBy('category', 'country')} will group all items of the same country, assuming that the \texttt{items} contain a property called \texttt{country}. We chose this API style to keep the number of public API methods small. The \texttt{splitBy} method supports the same proximity- and similarity-based subroutines for splitting piles. The \texttt{arrangeBy} offers coordinate-based (\texttt{xy}, \texttt{ij}, \texttt{uv}, and \texttt{index}) and data-driven (\texttt{data}) layouts, e.g., \texttt{arrangeBy('data', 'size')} will order items by a property called \texttt{country}.
Finally, the proximity-based \texttt{groupBy} subroutines can be re-evaluated automatically upon zooming, as the proximity between items might have changed (\autoref{fig:use-case-multiscale-navigation}). Similarly, the \texttt{arrangeBy} subroutines can be re-evaluated automatically after grouping piles.

\subsection{Adjust and Explore the Piling Interface via a GUI} \label{sec:pilingjs:gui}

\begin{figure}[t]
    \centering
    \includegraphics[width=1\columnwidth]{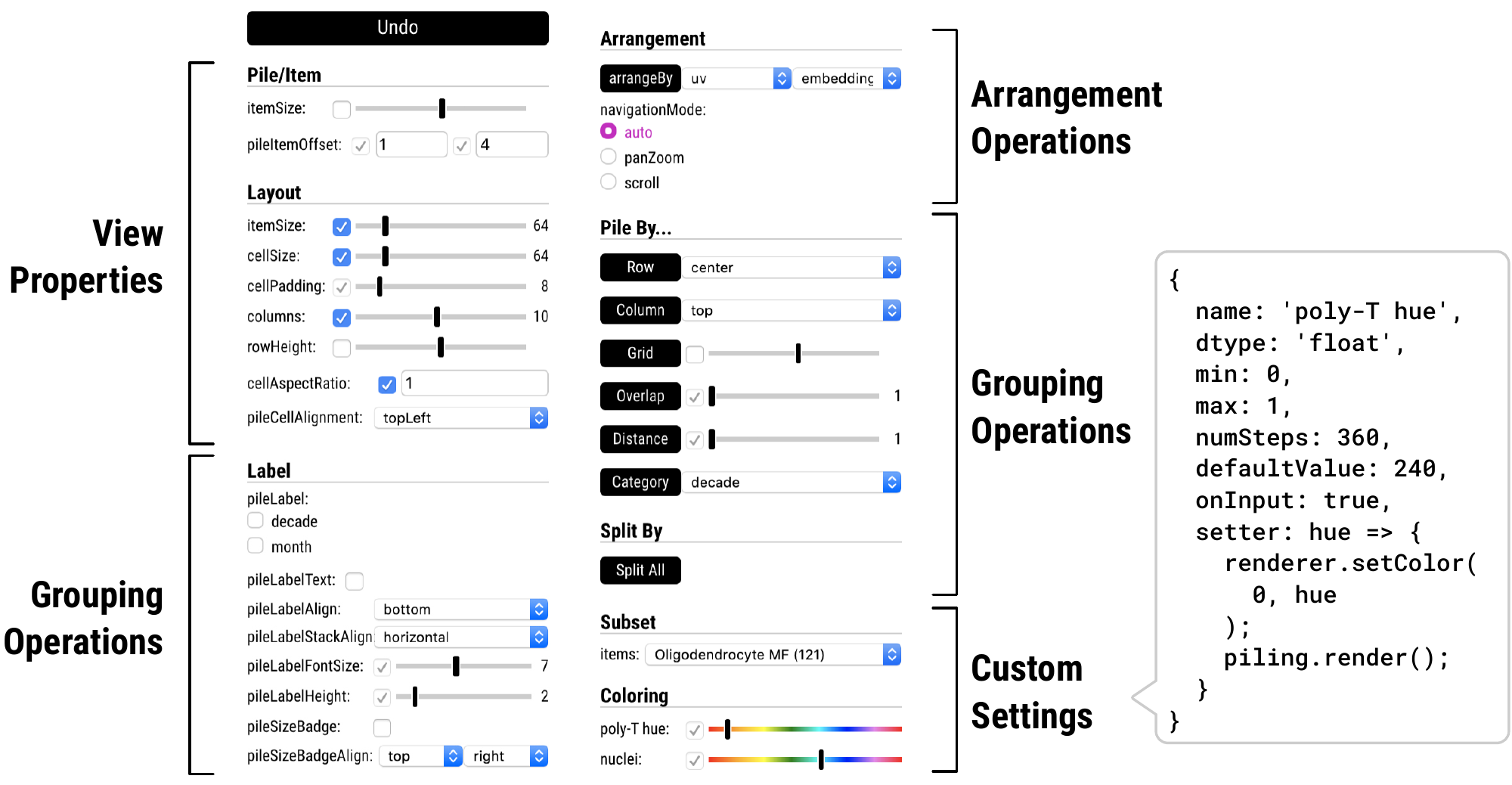}
    \caption{\textbf{Graphical User Interface for Parameterization.} \pilingjs implements several default settings and allows to define use-case specific settings. \new{See Supplementary Figure S4 for a scaled-up version.}}
    \label{fig:gui}
\end{figure}

\pilingjs provides a mid-level API, which hides the state and rendering aspects of visual piling but relies on the designer to implement the rendering pipeline programmatically. As we worked on the use cases (\autoref{sec:use-cases}), we realized that switching between a text editor and the browser to parameterize view specification can be time-consuming since the visual feedback is delayed until the browser refreshes. Therefore, as shown in \autoref{fig:gui}, we provide a simple yet effective GUI to allow the designer to adjust various view properties dynamically.

Currently, the GUI features elements for adjusting static property values such as Boolean flags, single or multiple selections, or numerical values. The GUI also has support for triggering groupings and arrangements. Finally, given the breadth of view properties, it is infeasible to cover every possible setting. Therefore, \pilingjs allows the designer to specify custom settings (\autoref{fig:gui} bottom-right).

\subsection{Implementation} \label{sec:pilingjs:implementation}

\pilingjs is implemented in JavaScript using PixiJS~\cite{pixijs} for WebGL rendering. We chose PixiJS for its highly-optimized 2D texture rendering and flexible mid-level API, which greatly simplifies the development of WebGL programs. The source code of \pilingjs is free and openly available at \url{https://github.com/flekschas/piling.js} and features extensive documentation (\url{https://piling.js.org/docs}) for all available view configurations.

\subsection{Performance Evaluation} \label{sec:pilingjs:performance-evaluation}

\begin{figure}[h]
    \centering
    \includegraphics[width=1\columnwidth]{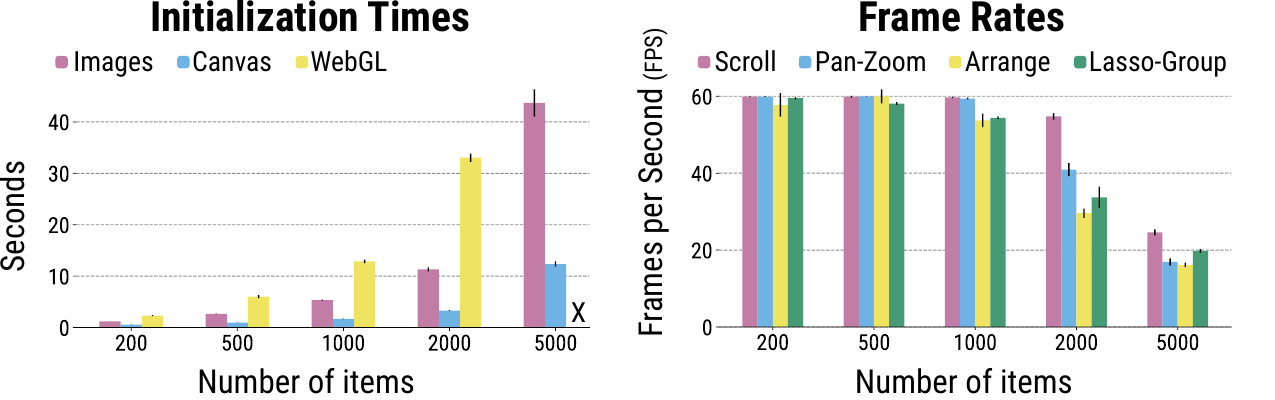}
    \caption{\textbf{Performance Evaluation.} Construction time (smaller is better) and frame rate (higher is better; 60 FPS is the best) as a function of the number of items for different media and interaction types. Loading 5,000 dynamic WebGL renderers (``X'') was not feasible as the browser timed out. Error bars show standard deviation.}
    \label{fig:performance-evaluation}
\end{figure}

\pilingjs is designed to support datasets of up to a few thousand items. In the following, we evaluate the initialization time and frame rate (\autoref{fig:performance-evaluation}). The initialization time includes the library's startup time, data rendering, and item creation. We compare the time for loading items of the following media types: images, canvas-derived textures, and WebGL programs. For the frame rate, we compare navigation, arrangement, and grouping animations \new{(i.e., animated transitions of the piling state triggered by scripted interactions)} using the example from~\autoref{fig:teaser}. Specifically, we examine scrolling, pan-and-zoom, automatic arrangements of all items, and lasso-based grouping, which cover the essential core interactions. We repeated each animation ten times and measured the duration in seconds and frames per second (FPS) in Chromium (v80) on a 2016 MacBook Pro.

As shown in \autoref{fig:performance-evaluation}, the initialization time increases with the number of items but remains acceptable until 1,000 items. The media type and size of the items greatly influences the initialization time. Especially the custom WebGL programs take longer to initialize. The frame rates for scrolling and pan-and-zoom interactions with datasets of up to 1,000 items is smooth but starts to degrade significantly for datasets larger than 2,000 items. As the arrangement and grouping animations are more involved, their frame rates are lower but remain acceptable for up to 2,000 items.

\section{Use Cases} \label{sec:use-cases}

In the following section, we present several use cases to demonstrate the generality of the visual piling approach and the expressiveness of our \pilingjs library for exploring large collections of small multiples. The use cases are also available online at \url{https://piling.js.org}.

\smallskip
\textbf{Compiling Training Data for Machine Learning.}

\begin{figure}[b]
    \centering
    \includegraphics[width=1\columnwidth]{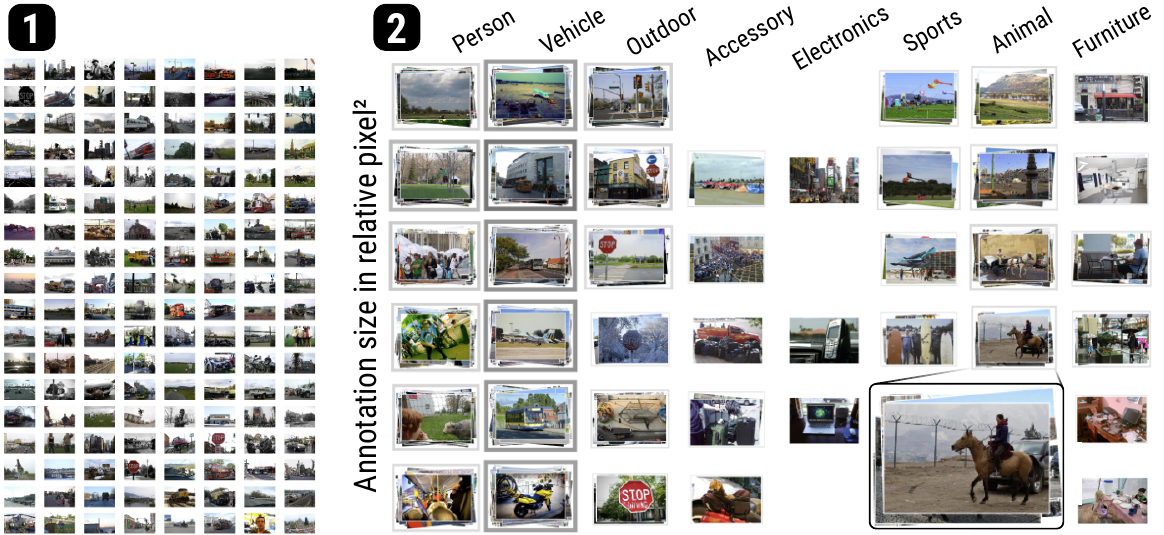}
    \caption{\textbf{Compiling Natural Images from the COCO dataset~\cite{lin2014microsoft}.} All images (1) contain car annotations, but only a subset of them show a car prominently. (2) Arranging the images by their primary annotation type and relative annotation size improves the explorability.}
    \label{fig:use-case-photos}
\end{figure}
A critical aspect of machine learning research, especially deep learning, is the composition of training and validation datasets to probe machine learning models. For instance, in computer vision research, this involves collecting, sorting, and selecting images. While several collections of annotated images exist for comparison and benchmarking~\cite{krizhevsky2009learning,deng2009imagenet,everingham2010pascal,lin2014microsoft}, subsets are often used during the initial development of the model for exploration. Creating these subsets is not trivial as nuanced image features might not have been extracted, and formal categorization of every potential interesting feature is prohibitive. Visual piling can address this issue by allowing users to sort existing datasets into subsets (\tgrouping) for rapid hypothesis testing. For example, in~\autoref{fig:use-case-photos}.1, we sampled 2\new{,}000 images of cars in context from the Microsoft COCO~\cite{lin2014microsoft} dataset that can be used to train car detector models. Browsing all images provides a first overview. Using the given object annotations, we can arrange (\tarrangement) the images in a two-dimensional grid by their primary category (x-axis) and relative size of the annotation in pixels\textsuperscript{2} (y-axis). Google Facets~\cite{googlefacets} allows for similar arrangements but requires zooming as the number of items increases. With \pilingjs we can instead group all items that are located within the same grid cell into piles, which provides visual cues about the groups' content and the ability to compose new groups manually.

\begin{figure}[t]
    \centering
    \includegraphics[width=1\columnwidth]{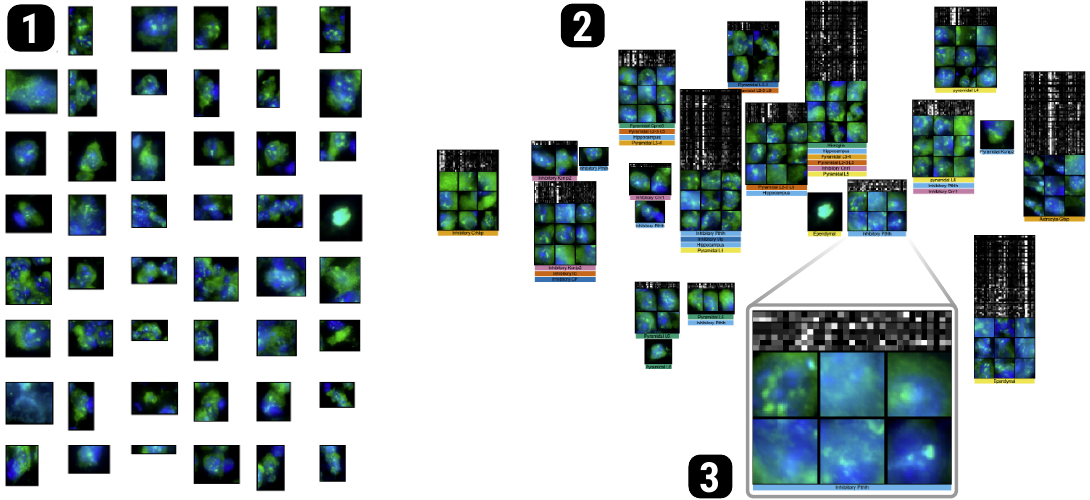}
    \caption{\textbf{Microscopy Cell Instances.} Small multiples of different cell types (1) from an immunofluorescence microscopy image~\cite{codeluppi2018spatial} where green shows total mRNA and blue represents nuclei. Visual piling (2) allows for simultaneous exploration of the cell phenotype and related gene expression profiles (black and white heatmap) as highlighted in (3).}
    \label{fig:use-case-microscopy}
\end{figure}

\smallskip
\textbf{Exploring Instance Annotations in Large Images.}
One aspect of analyzing large image data involves the exploration of instance annotations. For example, in cell biology, researchers annotate cell boundaries in immunofluorescence microscopy data of tissues or cell cultures. The goal of visual exploration is to compare and organize cells to each other for quality control and stratification (\tgrouping). Using a conventional small-multiples approach can be limiting when there are several potentially-interesting arrangements. In \autoref{fig:use-case-microscopy} we show an exploration of a microscopy image from Codeluppi et al.~\cite{codeluppi2018spatial}. Since the cells were clustered based on their gene expression profiles, we arranged cells by the gene expression data reported in the original paper (\tarrangement). As the cell bodies do not align well, we show a gallery preview of representative images (\autoref{fig:use-case-microscopy} 2 and 3) as the pile cover to highlight the diversity of cell images across the pile (\taggregation). Additionally, we preview individual cell annotations as one-dimensional heatmaps (\tpreviewing) above the cover, which show the cells' gene expression profiles. This enables us to correlate the cell morphology to the gene expression data.

\begin{figure}[b]
    \centering
    \includegraphics[width=1\columnwidth]{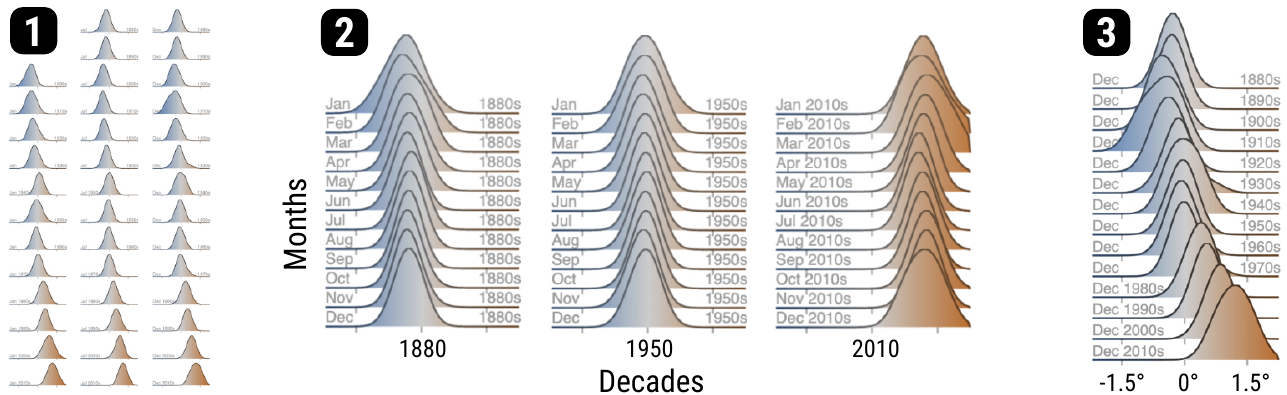}
    \caption{\textbf{Global Temperature Anomalies.} Surface temperature deviations from -1.5 to +1.5 degrees Celsius across decades. Piling enables the dynamic creation of ridge plot-like piles (2 and 3) from small multiples (1) for interactive comparison of decennial (2) and monthly (3) trends.}
    \label{fig:use-case-1d-repeated-measurements}
\end{figure}

\smallskip
\textbf{Comparing Repeated One-Dimensional Measurements.}
Comparing one-dimensional repeated measurements with small multiples typically involves the alignment of items along a shared axis to discover patterns (\tarrangement). For large numbers of repeated measurements, it can be beneficial to explicitly group measurements to emphasize trends \new{and to interactively change the grouping to highlight different patterns} between subsets of the data.
For example, in \autoref{fig:use-case-1d-repeated-measurements}, we loaded the global surface temperature anomaly dataset from NASA~\cite{lenssen2019improvements,gistemp}. This dataset contains surface temperature measurements for each month across 14 decades (the 1880s to 2010s) that is normalized by the mean temperature of 1951-1980. We plotted the mean temperature deviations from -1.5 to +1.5 degrees Celsius for each month of the 14 decades (\autoref{fig:use-case-1d-repeated-measurements}.1). Grouping the plots by decades or months, and arranging them by a vertical offset enables us to dynamically create ridge plot-like piles. Positioning the piles next to each other makes it easy to compare decennial (\autoref{fig:use-case-1d-repeated-measurements}.2) and monthly trends (\autoref{fig:use-case-1d-repeated-measurements}.3). We can now immediately see how the temperature increased over the last 140 years.

\begin{figure}[t]
    \centering
    \includegraphics[width=1\columnwidth]{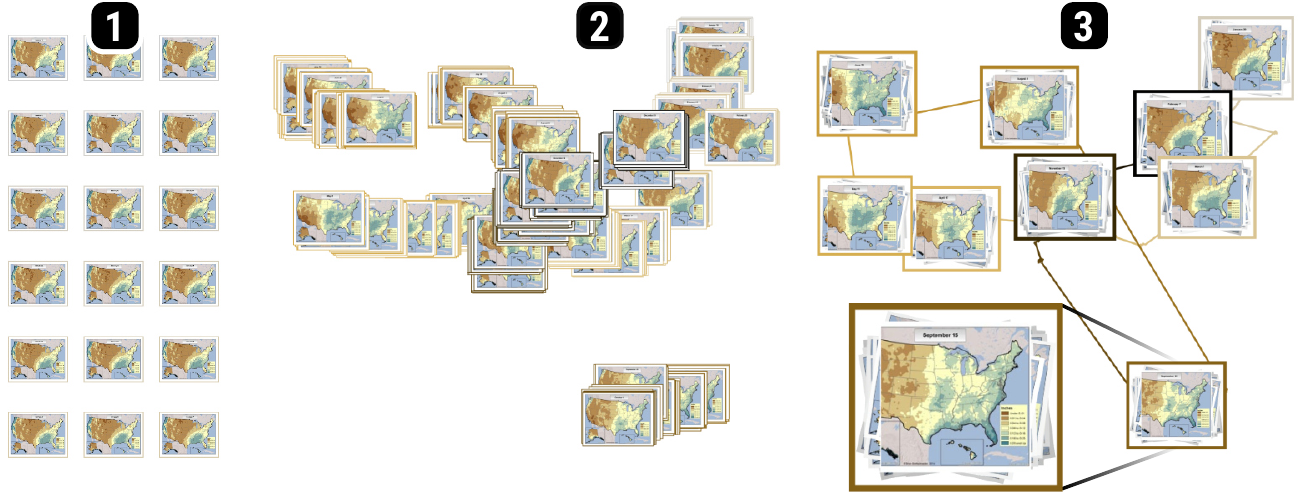}
    \caption{\textbf{Annual Precipitation Cycle.} Displaying movie frames as small multiples (1) does not uncover similar scenes well. After arranging the frames by their similarity (2) to show clusters and grouping the frames into piles, we can identify nine distinct states (3).}
    \label{fig:use-case-movie-analysis}
\end{figure}

\smallskip
\textbf{Movie Analysis.}
When analyzing movies, it can be insightful to study the visual similarity of scenes. To compare the similarity between frames, Bach et al.~\cite{bach2016time} folded a linear curve, called a time curve, in 2D space using a dimensionality reduction technique. In \autoref{fig:use-case-movie-analysis}.1, we loaded 365 frames from a movie showing the annual precipitation cycle of the United States~\cite{brettschneider2014normal} (one frame per day). Based on the similarity between each frame, we embedded the frames into a two-dimensional space with UMAP~\cite{mcinnes2018umap} (\autoref{fig:use-case-movie-analysis}.2). After arranging the frames by their embedding (\tarrangement), we can highlight the annual precipitation cycle and several \new{clusters} of highly similar frames. Visualizing the frames as thumbnails shows what these \new{clusters} represent. As the high number of frames makes it hard to compare individual frames, we grouped overlapping frames into piles to simplify the view (\tgrouping), which highlights nine visually distinct \new{precipitation patterns} (\autoref{fig:use-case-movie-analysis}.3). Additionally, we encode the frame order via the border color (which ranges from light gray (January) to black (December)) and connect the piles with a line visualization to foster the connection to the underlying sequence of the movie. \new{This line visualization is realized with D3~\cite{bostock2011d3} and linked to the pile interface.}

\begin{figure}[b]
    \centering
    \includegraphics[width=1\columnwidth]{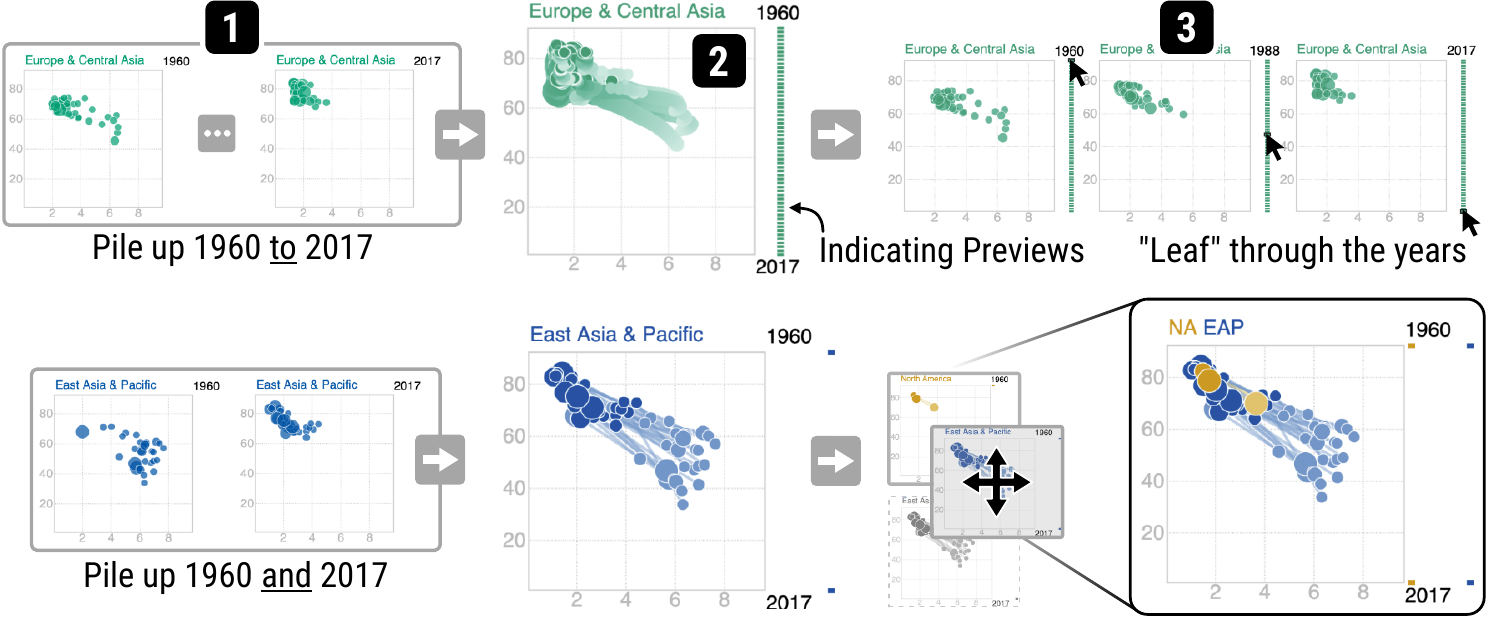}
    \caption{\textbf{Fertility Rate vs. Life Expectancy.} We group small multiples (1) into a pile (2) to compare changes over time, which we browse by leafing through the indicating previews (3). Manual grouping (bottom) allows us to compare East Asia (blue) against North America (yellow).}
    \label{fig:use-case-time-series}
\end{figure}

\smallskip
\textbf{Time Series Analysis.}
When dealing with time series, an important task is to identify overall trends and variations. To see and make sense of any trends, one must be able to compare individual items. Visual piling can address this challenge through content-aware browsing (\tbrowsing). In \autoref{fig:use-case-time-series}, we plot the fertility rate (x-axis) against life expectancy (y-axis) from Worldbank~\cite{worldbank} from 1960 to 2017 as small multiples, resolved by country and colored according to the geographic region. After grouping (\tgrouping) European countries (\autoref{fig:use-case-time-series}.2), we can see that, over time, the fertility rate lowers while the life expectancy increases, as shown by color gradient going from bright (1960) to dark (2017). To support comparing individual years without having to split the groups, we render small rectangles next to a pile as indicating previews (\tpreviewing) of the years. By leafing through the rectangles (\tbrowsing), the year's corresponding scatterplot is revealed, which allows us to trace the temporal development (\autoref{fig:use-case-time-series}.3 top). Upon manually grouping scatterplots, we aggregate (\taggregation) the data into a combined and connected scatterplot~\cite{haroz2015connected}, which is shown as the pile cover, to allow tracing the development of individual countries. By piling up the years 1960 and 2017 of North America and East Asia, we can see the alignment of countries in both regions (\autoref{fig:use-case-time-series}.3 bottom).

\begin{figure}[t]
    \centering
    \includegraphics[width=1\columnwidth]{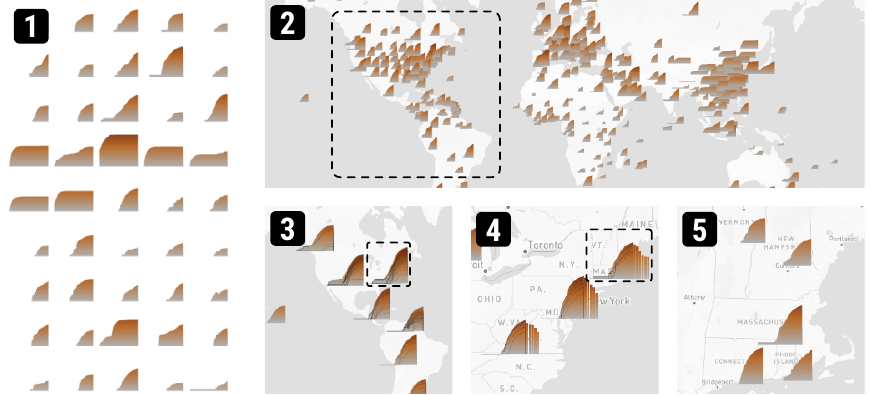}
    \caption{\textbf{Worldmap of COVID-19 Infection Rates.} Small multiples of area charts (1) show the number of infected people over time. Arranging the charts geographically (2) and grouping them by overlap (3) highlights infection hot spots without overplotting issues. Upon zooming in, piles are automatically split (3-5).}
    \label{fig:use-case-multiscale-navigation}
\end{figure}

\smallskip
\textbf{Pattern-Driven Navigation in Multiscale Visualization.}
A common challenge in exploring local patterns in multiscale visualization is the lack of visual details at an overview~\cite{jul1998critical}. These details are often needed to decide which region to explore in detail. Lens techniques can be applied to magnify a selected region, but many lens techniques do not scale well to large numbers of local patterns. As shown in Scalable Insets~\cite{lekschas2019pattern}, the scalability issue can be addressed by displaying local patterns as insets and grouping those insets into piles upon zooming out. In \autoref{fig:use-case-multiscale-navigation}.1, we show area charts of COVID-19 infection rates as small multiples. By arranging the small multiples according to their geolocation (\tarrangement) we can gain an overview of the global spread (\autoref{fig:use-case-multiscale-navigation}.2). To avoid issues of overplotting, we group (\tgrouping) overlapping items into piles (\autoref{fig:use-case-multiscale-navigation}.3). Piles are visually represented by a stacked area chart to show the overall regional spread of the virus (\taggregation). The combination of grouping and aggregating provides guidance without introducing severe occlusion. Browsing individual countries (\tbrowsing), states, or counties is realized by navigating to a specific area. Upon zooming in, piles are automatically split when the items do not overlap anymore, given their original position (\autoref{fig:use-case-multiscale-navigation}.4).

\begin{figure}[t]
    \centering
    \includegraphics[width=1\columnwidth]{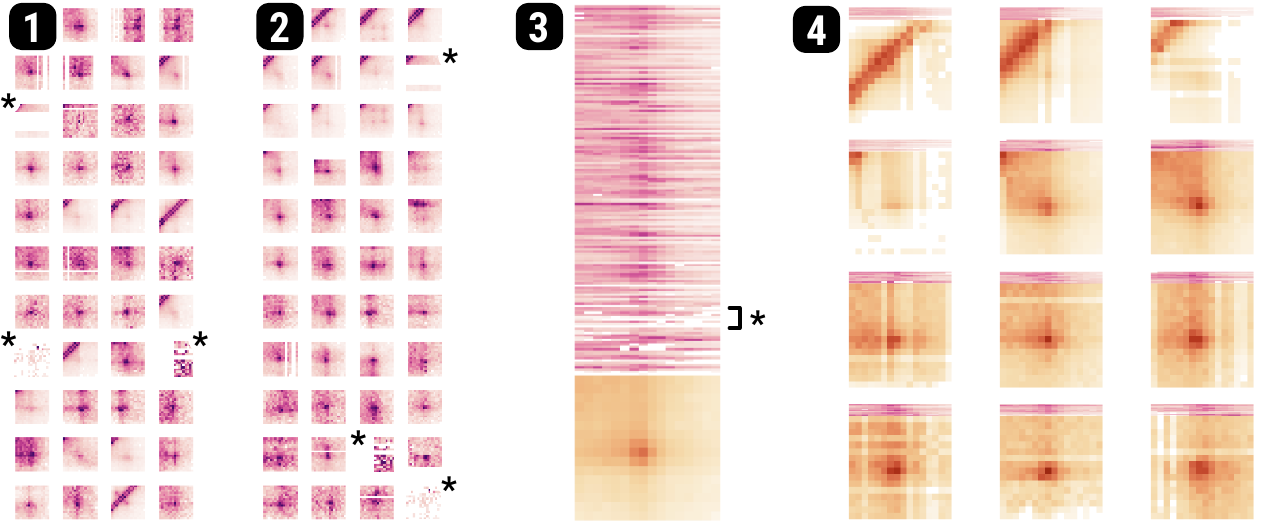}
    \caption{\textbf{Comparison of Matrix Patterns.} Small multiples of matrix patterns (1) that are supposed to show a dark dot in the center. Ordering (2), aggregating (3), and grouping (4) through visual piling enables us to discover overall trends and outliers (*).}
    \label{fig:use-case-matrix-patterns}
\end{figure}

\smallskip
\textbf{Matrix Pattern Comparison.}
A common task in analyzing network data involves the detection and assessment of reoccurring pattern instances, known as motifs. When the data is visualized as a matrix, these motifs can be represented as small multiples. In analyzing motifs, single instances provide only limited insight. Instead, the analyst typically needs to compare individual motifs to groups of motifs. Using a piling interface, this comparison can be achieved by interactively grouping and aggregating the patterns into piles. For example, \autoref{fig:use-case-matrix-patterns} shows pattern instances from Rao et al.~\cite{rao20143d}, which should show a dark dot in the center and act as proxies for specific biological events. As these instances are retrieved computationally, the goal is to verify if the expected pattern is truly exhibited. Scanning over the small multiples sequentially is time-consuming but highlights differences (\autoref{fig:use-case-matrix-patterns}.1). Ordering the small multiples (\autoref{fig:use-case-matrix-patterns}.2) helps to find instances with the expected pattern (\tarrangement). By aggregating (\taggregation) all instances and showing the average as the pile cover, we can confirm that, on average, the algorithm works as desired (\autoref{fig:use-case-matrix-patterns}.3). However, by additionally showing one-dimensional previews on top of the cover, we can identify many outliers, which should be removed prior to subsequent analyses (\autoref{fig:use-case-matrix-patterns}.3 asterisk). Interactive grouping also enables us to stratify (\tgrouping) the pattern collection for more efficient data cleaning (\autoref{fig:use-case-matrix-patterns}.4).

\section{Discussion \& Future Work} \label{sec:discussion}

The piling metaphor is a generic and flexible approach for organizing and exploring small multiples. We have demonstrated a variety of use cases, but visual piling can be useful for any small-multiple setup that benefits from interactive re-organization and comparison of the items. This is, for instance, the case when different arrangements can be insightful or multiple aggregated items are to be compared. \new{However, to drive new insights, visual piling requires at least some user interaction.}
With the development of \pilingjs our goal is to provide a domain- and application-independent toolkit that is extensible to many different use cases. To this end, we decoupled the rendering and aggregation methods in \pilingjs to allow the designer to provide their own solutions.
\new{Depending on the data and visualization type of the items}, the visual piling approach can potentially scale to tens of thousands of items given the application of scalable arrangement, browsing, and aggregation methods.
\new{For example, to visually scale to piles with many items, the designer needs to choose a previewing layout that either naturally limits the number of previewed items, e.g., partial, gallery, or combining) or applies a preview aggregator that limits the number of previewed items.}
Technically, to improve the initialization time of \pilingjs for large datasets, we are planning to bundle multiple items into larger textures first and upload the textures onto the GPU in batches. To address the rendering performance, we will optimize the interaction handling. As non-interactive rendering scales to 10,000 and more items, dynamically enabling and disabling interaction handling could greatly improve the performance.

\new{The intended target users of \pilingjs are visualization designers who are familiar with web development. While \pilingjs greatly simplifies the creation of piling interfaces, the user still needs to understand and wrangle the source data to design an effective visualization for the individual small multiples. From our experience in implementing the eight use cases, we learned that designing a piling interface with \pilingjs requires a holistic approach during the development of the different design dimensions, as the dimensions are not independent.}

Our structured design space definition provides a framework to \new{guide the} design of new interactive visual piling interfaces. In combination with our \pilingjs toolkit, we hope that this work will increase the accessibility of visual piling for a wide range of use cases. Beyond the application of piling to new domains, one exciting direction for future work is the evaluation of the perceptual effectiveness of different previewing approaches. Similarly, many different pen, touch, and mouse gestures can be employed to trigger grouping, arrangement, and browsing interactions. We hope that \pilingjs will streamline user studies development to understand the performance of different piling techniques better.

\acknowledgments{
We would like to thank Spandan Madan for his insightful discussions and ideas regarding the machine learning use case. This work was supported in part by the National Institutes of Health (U01 CA200059 and OT2 OD026677).}

\bibliographystyle{abbrv-doi}
\bibliography{___references}

\end{document}


\begin{titlepage}
 \begin{center}
 {\fontfamily{ptm}\selectfont
  {\huge \textbf{A Generic Framework and Library for}}\\[.33cm]
  {\huge \textbf{Exploration of Small Multiples Through}}\\[.33cm]
  {\huge \textbf{Interactive Piling}}\\[1.25cm]
  {\LARGE \textbf{Supplementary Material}\\[1.75cm]}

  {\Large Fritz Lekschas}\\
  {Harvard School of Engineering and Applied Sciences}\\[.5cm]

  {\Large Xinyi Zhou}\\
  {State Key Lab of CAD\&CG, Zhejiang University}\\[.5cm]

  {\Large Wei Chen}\\
  {State Key Lab of CAD\&CG, Zhejiang University}\\[.5cm]
  
  {\Large Nils Gehlenborg}\\
  {Harvard Medical School}\\[.5cm]
  
  {\Large Benjamin Bach}\\
  {University of Edinburgh}\\[.5cm]

  {\Large Hanspeter Pfister}\\
  {Harvard School of Engineering and Applied Sciences}\\[1.75cm]

 }
 \end{center}
 \end{titlepage}

\begin{figure}[h!]
    \centering
    \includegraphics[width=1\textwidth]{teaser-extension}
    \caption{\textbf{Grouping and Arrangement Refinement From the Teaser Figure.} After grouping items in close proximity into piles in Fig. 1A3, we further refine the grouping and isolate four piles via manual rearrangement to show overarching concepts in how people think of necklaces. Please also see our Supplementary Video for a recorded exploration of this dataset.}
    \label{sfig:teaser-extension}
\end{figure}

\clearpage

\begin{figure}[h!]
    \centering
    \includegraphics[width=0.95\textwidth]{data-view-spec}
    \caption{\textbf{View Specification.} We exemplify how the declarative view specification (bold source code) enables different pile encodings. In (1), we shows the default pile encoding. In (2), we randomized the item offset for partial previewing. In (3), we implemented foreshortened previews. And in (4), we visualize the pile as a gallery preview.}
    \label{sfig:data-view-spec}
\end{figure}

\clearpage

\begin{figure}[h!]
    \centering
    \includegraphics[width=0.9\textwidth]{pilingjs-arrange-group-by}
    \caption{\textbf{Arrangement and Grouping in \pilingjs.} We demonstrate how \pilingjs' \texttt{arrangeBy} and \texttt{groupBy} methods work. (Top) The \texttt{arrangeBy('data')} subroutine expects an array of item property names (e.g., \texttt{distanceToDiagonal}, \texttt{noise}, and \texttt{size}) to arrange the piles by one, two, or more dimensions. Multidimensional cluster plots are realized with UMAP. (Bottom) The \texttt{groupBy} can take the current layout  (e.g., grouping by \texttt{overlap}), item properties (e.g., grouping by \texttt{cellSubType} \texttt{category}), or similarity (e.g., grouping by \texttt{cluster}) into consideration. The \texttt{cluster} subroutine uses k-means clustering with $k = max(2, \ceil{\sqrt{\vert\mathit{items}\vert / 2}})$ by default.}
    \label{sfig:pilingjs-arrange-group-by}
\end{figure}

\clearpage

\begin{figure}[h!]
    \centering
    \includegraphics[width=1\textwidth]{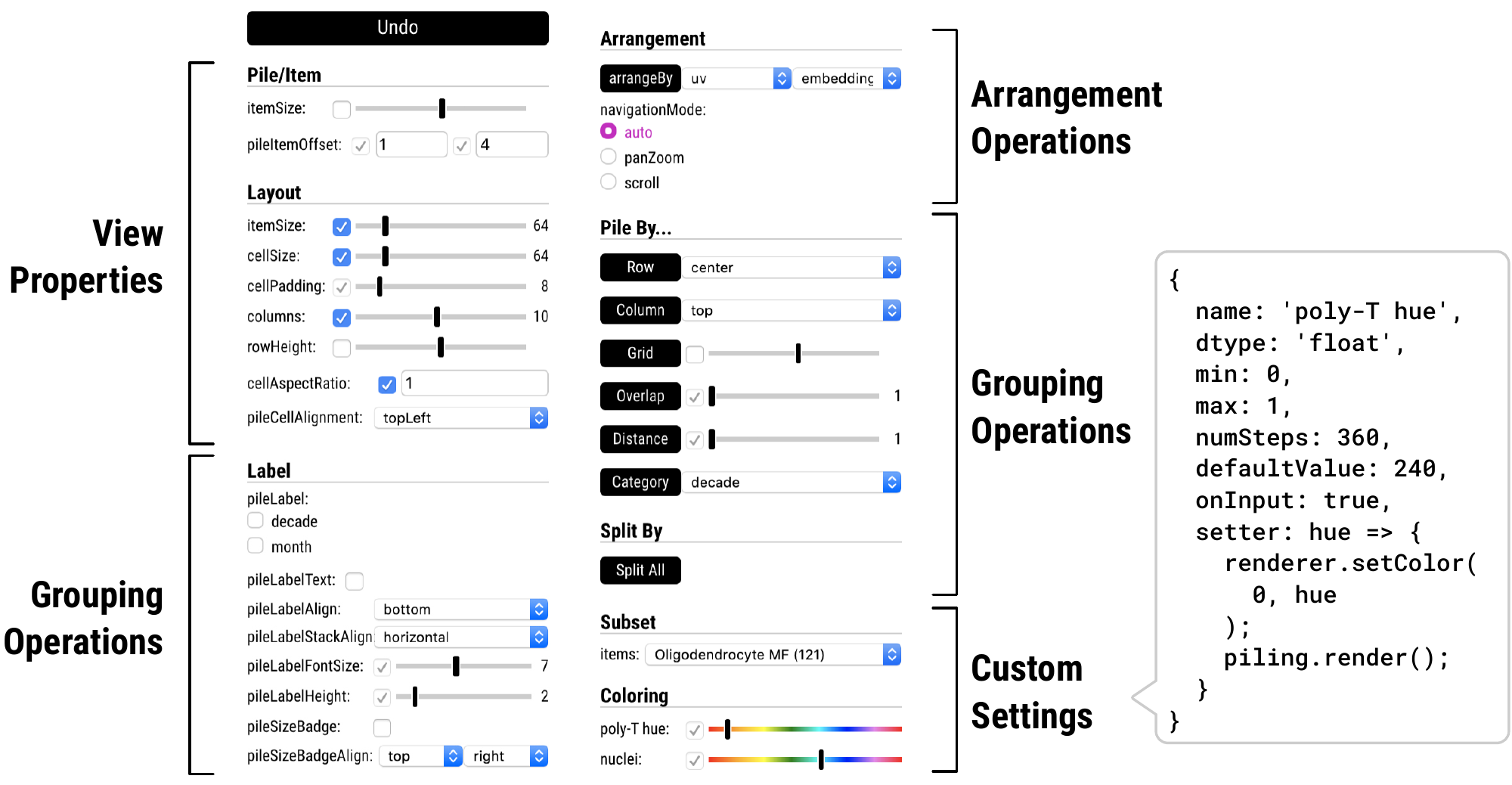}
    \caption{\textbf{Graphical User Interface for Parameterization.} \pilingjs implements several default settings and allows to define custom use-case specific settings. \emph{(Scaled-up version of Fig. 10.)}}
    \label{sfig:gui}
\end{figure}

\clearpage

\begin{table}[h]
    \caption{\textbf{Piling.js Coverage of the Visual Piling Design Space.} An overview of piling.js' current support for the five dimensions of the visual piling design space.}
    \setlength\tabcolsep{5pt}
    \renewcommand{\arraystretch}{1.0}
    \begin{tabularx}{\columnwidth}{lX}
        \toprule
        \multicolumn{2}{c}{\textbf{Grouping}} \\
        \midrule
        Manual & Sequential grouping via a drag \& drop gesture \\
               & Sequential grouping via multi-selections (multiple mouse clicks while holding down the \texttt{shift} key) \\
               & Parallel grouping via a lasso gesture \\
        Automatic & Layout-driven grouping via \small{\texttt{groupBy('row')}}, \small{\texttt{groupBy('column')}}, or \small{\texttt{groupBy('grid')}} \\
               & Proximity-based grouping via \small{\texttt{groupBy('overlap')}} or \small{\texttt{groupBy('distance')}} \\
               & Similarity-based grouping via \small{\texttt{groupBy('category')}} or \small{\texttt{groupBy('cluster')}} \\
        \midrule
        \multicolumn{2}{c}{\textbf{Arrangement}} \\
        \midrule
        Item & Random offset arrangement \\
             & Orderly rule-based arrangement \\
             & Orderly data-driven arrangement \\
        Pile & Gridded linear ordering arrangement via \small{\texttt{arrangeBy('index')}} \\
             & Gridded 2D spreadsheet-like arrangement via \small{\texttt{arrangeBy('ij')}} \\
             & Precise arrangement via \small{\texttt{arrangeBy('xy')}} or  \small{\texttt{arrangeBy('uv')}} \\
             & Precise data-driven arrangement via \small{\texttt{arrangeBy('data')}} \\
        \midrule
        \multicolumn{2}{c}{\textbf{Previewing}} \\
        \midrule
        Partial & Preview via partial item overlap \\
        Gallery & Preview gallery of 2, 3, 4, 6, 8, or 9 thumbnails \\
        Foreshortened & Preview via aggregation or a custom preview renderer \\
        Combining & Preview via aggregation or a custom cover renderer \\
        Indicating & Preview via custom preview renderer \\
        \midrule
        \multicolumn{2}{c}{\textbf{Browsing}} \\
        \midrule
        In-place & Full item will appear at the top of the pile upon hovering the associated preview \\
        Dispersive & Temporary dispersion into a regular 1D or 2D grid \\
        Layered & Browse a pile in isolation via the context menu \\
        Hierarchical & Nested browsing of sub-piles in isolation via the context menu \\
        \midrule
        \multicolumn{2}{c}{\textbf{Aggregation}} \\
        \midrule
        Synthetic & Summary statistics (e.g., min, max, mean, median, or sum) \\
                  & Custom aggregator with a preview or cover renderer \\
        Representative & Kmeans cluster centroids \\
                       & Custom aggregator \\
        Simplistic & Custom aggregator with custom preview or cover renderer \\
        \bottomrule
    \end{tabularx}
    \label{stab:pilingjs-implementation-coverage}
\end{table}